\documentclass[times,twocolumn]{article}

\usepackage{simpleConference}
\usepackage{framed,multirow}

\usepackage{amssymb}
\usepackage{latexsym}

\usepackage{url}
\usepackage{xcolor}
\definecolor{newcolor}{rgb}{.8,.349,.1}

\usepackage{hyperref}

\usepackage[switch,pagewise]{lineno} 

\usepackage{graphicx}
 
\usepackage{multirow}
\usepackage{subcaption}
\usepackage{booktabs}
\usepackage{amsmath, amsfonts}
\usepackage[normalem]{ulem}
\usepackage{booktabs}
\usepackage{tikz}
\usetikzlibrary{calc,spy}
\usepackage{nicefrac}
\usepackage{wrapfig}
\usepackage{enumitem}

\usepackage{authblk}

\makeatletter
\renewcommand\AB@affilsepx{, \protect\Affilfont}
\makeatother

\author[1]{Beatrix-Emőke Fülöp-Balogh}
\author[2]{Eleanor Tursman}
\author[2]{James Tompkin}
\author[1,3]{Julie Digne}
\author[1,3]{Nicolas Bonneel}
 
\affil[1]{Univ Lyon}
\affil[2]{Brown University}
\affil[3]{CNRS}

\begin{document}

\title{\large{Dynamic Scene Novel View Synthesis via Deferred Spatio-temporal Consistency}}%

\maketitle
\thispagestyle{empty}

\begin{abstract}
Structure from motion (SfM) enables us to reconstruct a scene via casual capture from cameras at different viewpoints, and novel view synthesis (NVS) allows us to render a captured scene from a new viewpoint. Both are hard with casual capture and dynamic scenes: SfM produces noisy and spatio-temporally sparse reconstructed point clouds, resulting in NVS with spatio-temporally inconsistent effects.
We consider SfM and NVS parts together to ease the challenge. First, for SfM, we recover stable camera poses, then we \emph{defer} the requirement for temporally-consistent points across the scene and reconstruct only a sparse point cloud per timestep that is noisy in space-time. Second, for NVS, we present a variational diffusion formulation on depths and colors that lets us robustly cope with the noise by enforcing spatio-temporal consistency via per-pixel reprojection weights derived from the input views.
Together, this deferred approach generates novel views for dynamic scenes without requiring challenging spatio-temporally consistent reconstructions nor training complex models on large datasets. We demonstrate our algorithm on real-world dynamic scenes against classic and more recent learning-based baseline approaches.
\end{abstract}

\vspace{-1em}
\begin{figure*}
\centering
    \includegraphics[scale=0.8]{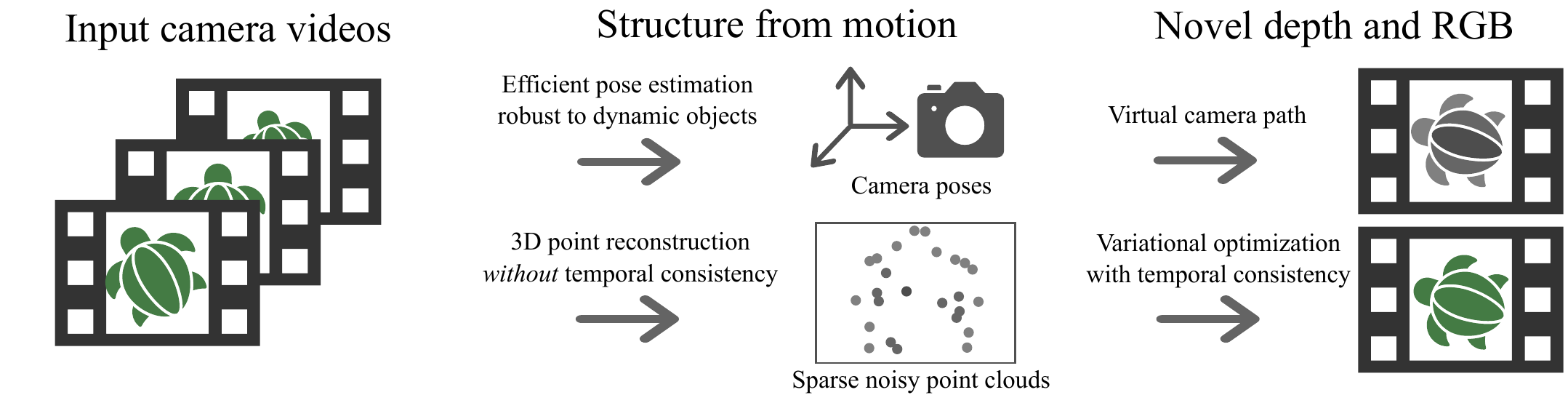}
    \vspace{-0.25cm}
    \caption{
    Given a small set of video sequences of a performance, our method computes camera poses and sparse points, then optimizes those points into a novel video sequence following a user-defined camera path. Our space-time SfM intentionally does not compute temporal consistency for points on dynamic objects and instead defers spatio-temporal consistency in both depth and RGB reconstructions to the novel view synthesis stage via our variational formulation.
    }
    \label{fig:architecture}
\end{figure*}
\section{Introduction}
Novel-view synthesis (NVS) creates a new view of a scene by combining existing images captured from different viewpoints. Much progress in NVS has been made over the past two decades to tackle its two core problems: 1) how to build a proxy scene geometry to aid in rendering, such as constructing simplified sparse depth points or a piecewise planar mesh via structure from motion (SfM), and 2) how to interpolate or extrapolate an image via the reprojected proxy given the existing captured imagery. NVS increases in difficulty across many axes: as the cameras become farther apart (wide baseline), as their number decreases (few camera), as they become handheld (casual capture), as the scene itself contains motion (dynamic scene), as the scene phenomena become more visually complex (geometry, materials, and motion), and as the time given to generate the result decreases (compute cost).

We consider dynamic scenes captured by a small number of cameras (5--12) over baselines of around 60$^\circ$, as might occur with a crowd of people capturing an event (Figure~\ref{fig:architecture}). 
Within this scenario, we include sequences with casual handheld cameras. This is a relatively rare and challenging setting because both the cameras and the scene objects move simultaneously, and because sequences with only a small number of casual cameras makes robustness hard to obtain.
This complicates camera pose estimation and depth estimation in SfM and, if the proxy geometry is not perfect, causes ghosting, bleeding, and flickering artifacts across views and time during NVS in both moving objects and the background. 
Thus, one key component of any algorithm is a way to enforce spatio-temporal consistency in both the SfM and the NVS to reduce these artifacts. 


We propose to address these challenges by \emph{deferring} the difficult problem of reconstructing dynamic objects in time via SfM, and instead using a NVS approach to enforce temporal consistency. To ease the task of reconstructing dynamic scenes via SfM, many approaches first segment out moving objects or feature points and process the static background and the dynamic foreground separately \cite{vo2016spatiotemporal, Mustafa16,Mustafa19}. Instead, we first recover camera poses for all views without any explicit dynamic object segmentation. Then, we recover scene points on both static and dynamic objects without temporal consistency and performing per-frame SfM across views only. This is easier to solve, but leads to significantly noisy reconstructions temporally.

%

Next, we turn our sparse (and noisy) reconstructed point clouds into novel views. This is commonly completed by densifying points into a depth map~\cite{holynski2018fast} for each view in a consistent way, and using the depths to reproject and merge input RGB views into a virtual view. We present a formulation which only densifies a depth map in the virtual camera's view, rather than for all input views, which leads to a more efficient solve. For this, we take a coarse-to-fine variational approach and solve a diffusion-based formulation. Importantly, this formulation lets us enforce robust temporal consistency in the output depth to overcome the initial noisy reconstructions from the SfM. To determine our final RGB values, we also solve for the output color within the coarse-to-fine variational formulation.

We perform comparisons to recently-proposed approaches in point densification and view interpolation, using both optimization and learning-based approaches. Further, we show results on a synthetic dataset in an ablation study. In a nutshell, we show that considering SfM and NVS together allows us to ease the difficult temporally-consistent reconstruction problem and instead cope with it at the rendering stage. Overall, our work takes another step forward in improving digital content creation for scenes captured by multiple video cameras.

\section{Related Work}
Rendering a novel viewpoint of a real-world scene captured with photographs is a problem that has received much attention over the past 30 years~\cite{zhang2004survey}.

 
\paragraph*{Static scene IBR}
Image-based rendering (IBR) has initially attempted to render static scenes either from set of images or videos. This can be achieved either via warping input views using optical flow~\cite{chen1993view}, using coarse geometric proxies~\cite{gortler1996lumigraph} or via deep learning approaches~\cite{flynn2019deepview}.
In complex environments, IBR techniques often need some 3D proxy reconstruction. For example, the Lumigraph~\cite{gortler1996lumigraph,buehler2001unstructured} uses planar or coarse geometric proxies; Shade et al.~\cite{shade1998layered} used multiple planar sprites; and Debevec et al.~\cite{debevec1996modeling} employed photogrammetric reconstructions of buildings. 
Others have used 3D meshes from multi-view stereo reconstructions~\cite{snavely2006photo,hedman2016scalable}. 
For instance, Chaurasia et al.~\cite{chaurasia2013depth} proposed a depth-based synthesis using planar superpixel patches~\cite{achanta2010slic}. Matzen et al.~\cite{Matzen17} used two spherical cameras to synthesize an omni-directional stereo panorama.
Recently, Riegler and Koltun~\cite{Riegler2020FVS} synthesized new views via neural textures atop a Delaunay reconstruction of sparse points obtained from video of static scenes.
Beyond surface geometry, NeRF~\cite{nerf} performs an expensive optimization to create a volumetric function that is then rendered to synthesize new views.

Solving problems in the gradient domain can help too; for instance, to achieve smoother interpolations~\cite{kopf2013image} or to densify sparse scene points. Holynski and Kopf spatio-temporally propagate sparse depth samples in a single view by solving a Poisson problem~\cite{holynski2018fast}. This method relies on camera motion to detect depth edges, which limits it to static scenes.
Inspired by gradient domain approaches, we formulate a variational approach that jointly enforces depth smoothness and consistency, color smoothness and consistency, as well as temporal consistency.
Our approach additionally works with multiple potentially-dynamic cameras, and introduces a view-consistency term to ensure geometric consistency between views.

Deep learning can also be employed for static scene IBR. This includes plane sweep volumes~\cite{flynn2016deepstereo} and multi-plane images to interpolate between two static narrow-baseline views~\cite{zhou2018stereo} or between multiple views at once~\cite{llff,flynn2019deepview}, appearance flows to generate novel views from a single image of isolated objects~\cite{zhou2016view}, and light-field view interpolation~\cite{kalantari2016learning}. 
Hedman et al.~\cite{deepblending} use a geometric proxy and learn blending weights between view reprojections using a CNN. To improve the quality around depth discontinuities, Choi et al.~\cite{evs} use a 3D uncertainty volume as a proxy and neural network-based patch refinement.
Srinivasa et al.~\cite{Srinivasan17} train a CNN to predict a light field from a single image for small-baseline view synthesis. Similarly, Song et al.~\cite{song19} synthesize new views from a single image of a static scene using deep learning.

While these techniques were not designed for videos and so neither explicitly maintain temporal consistency nor are constrained by speed, we nevertheless compare our approach to relevant methods for static scenes taken frame by frame.




\paragraph*{Dynamic scene VBR}


For dynamic scenes, please see dos Anjos et al.~\cite{dos2018navigation} for an exhaustive survey on video-based rendering (VBR) techniques. 
%
%
The need for controlled capture setting is shared by many methods. Zitnick et al~\cite{Zitnick2004} use a specific system of 8 cameras combined with segmentation based stereo to extract the geometry.
Similarly Wilburn et al.~\cite{Wilburn05} use an array of 100 tightly-packed cameras. Broxton et al.~\cite{broxton2020immersive} describe a custom camera array of $46$ synchronized cameras mounted on a dome used to capture 6DoF wide-baseline light field videos. Guo et al.~\cite{relightables} relight video with a set up of $331$ light sources and $90$ cameras, while Collet et al.~\cite{Collet15} require $106$ cameras.
In a less constrained way, Pozo et al.~\cite{Pozo16} create a 16-camera rig to reconstruct 360 panoramic videos and synthesize new views. 
Penner and Zhang~\cite{penner2017soft} use a soft volumetric representation for narrow baseline IBR to enforce smooth reconstructions. This method can handle motion, but has trouble handling unstructured data and works best from camera arrays. Our method also works with handheld cameras.

Casually-captured videos have also been considered. Ballan et al.~\cite{ballan2010unstructured} allow for quick transitions between handheld video sequences. Their method segments a single dynamic foreground subject approximated by a planar proxy, and creates a 3D reconstructed static background. To cope with dynamic background objects reprojecting incorrectly, the method blurs background transitions between captured viewpoints. Our method assumes no segmentation nor planarity assumptions for dynamic objects.
Lipski et al.~\cite{Lipski10cgf} use dense correspondence fields to interpolate views between videos. They disambiguate matches in difficult cases by manually drawing correspondence lines on image pairs to use as priors in their matching algorithm.
Mustafa et al.~\cite{Mustafa16, Mustafa19} reconstruct isolated moving objects after segmenting them out from the initial video. These methods focus on specific object meshes, and so do not provide re-rendering of an entire scene from a novel viewpoint.

Recently, Luo et al.~\cite{Luo-VideoDepth-2020} introduced a consistency term by fine tuning a neural network to improve the estimated depth per point. This works for a single camera with no or limited dynamic motion. Bansal et al.~\cite{bansal20} use foreground and background extraction together with a self-supervised CNN based composition operator, and Yoon et al.~\cite{yoon20} use deep learning to extrapolate new views from a single monocular video camera; we compare our approach to this method.

Outside of NVS, other video reconstruction tasks raise consistency questions. Vo et al.~\cite{vo2016spatiotemporal} used a spatio-temporal bundle adjustment technique and human motion priors to reconstruct actor performances by temporally aligning videos at sub-frame precision. Bao et al.~\cite{bao2019depth} using deep learning for consistent video super resolution. Finally, Davis et al.~\cite{davis2003spacetime} recovered depth in dynamic scenes by unifying structured light and laser scanning into a space-time stereo framework.

\section{Method}
Our algorithm takes as input a set of casually-captured synchronized videos. We also provide the focal lengths for a pair of cameras (required by OpenMVG~\cite{moulon2016openmvg}), while the remaining focal lengths are estimated automatically by our algorithm. Our method proceeds in two steps (Figure \ref{fig:architecture}):

\begin{enumerate}
    \item \textbf{Camera pose estimation and 3D scene points.} We perform a three-step structure from motion reconstruction to provide both the set of camera poses and a set of sparse 3D points for each time step (Section \ref{sec:pose_estimate}).
    \item \textbf{Novel depth and novel view rendering.} We densify the sparse points into a depth map and render a new virtual camera frame by optimizing a coarse-to-fine variational formulation while enforcing spatio-temporal consistency (Section~\ref{sec:rendering}).
\end{enumerate}

\subsection{Camera pose estimation and 3D scene points.}
\label{sec:pose_estimate}

Let us consider a set of $S$ synchronized video views of a dynamic scene, each composed of $T$ frames. We call $I=\{I_{s,t}|s=1,..., S; t=1,...,T\}$ the set of all frames indexed by $s$ (camera index) and $t$ (time step). 
At each frame, via SfM, we recover the camera parameters $\mathbf{C}_{s,t}$ consisting of the intrinsic matrix and extrinsic rotation and translation matrices, and a set of sparse 3D points for each time step. First, we efficiently recover a set of camera poses for all frames. In contrast to other methods~\cite{ballan2010unstructured,Mustafa16,Mustafa19}, we estimate poses without an explicit dynamic object segmentation step. Second, we recover 3D points by solving a per-timestep SfM problem without a complex temporal reconstruction.
We solve each SfM problem with an \emph{a contrario} algorithm~\cite{Moulon2012}. This automatically adapts thresholds to the input data instead of using global thresholds, which is more flexible to different inputs.

\paragraph*{Efficient camera pose estimation}
A straightforward approach for accurate SfM is to solve a problem across all frames simultaneously, but this can be expensive and memory prohibitive. A second approach might consider solving only between consecutive time steps, but this is known to produce camera position drift \cite{Cornelis04}. Instead, we take a coarse-to-fine approach.

We begin by computing SfM across keyframes at every $\kappa$ time steps of each video. We detect and match SIFT keypoints within this subset and then simultaneously solve for all camera poses and 3D points. %
Then, we refine our estimate with a second SfM that only matches keypoints between successive frames of the same camera view, with previously-estimated camera poses held fixed. This considers every frame of every video, but we only match $I_{s,t}$ to $I_{s,t+1}$, and not to $I_{s+1,t}$ or $I_{s+1,t+1}$. To recover smooth camera paths per view, we add two additional penalty terms to the bundle adjustment: 
\begin{align}
\label{eq:camera_center_constraint}
 w(t-t')\ \|C_{s,t} - C_{s,t'}\|^2, ~~~~ t - 3 \leq t'\leq t + 3
\end{align}
and
\begin{align}
\label{eq:camera_center_constraint2}
 w(t-t')\ \|A_{s,t} - A_{s,t'}\|^2, ~~~~t - 3 \leq t'\leq t + 3,
\end{align}
where $w(t-t')$ is a Gaussian weight function, $C_{s,t}$ is the center of each camera pose, and $A_{s,t}$ is the angle-axis representation of the rotation matrix $R_{s,t}$.
This second SfM reduces computation time over all-pairs matching while still reducing drift by constraining the frame-to-frame pose estimates by the keyframe pose estimates.
For hyperparameters, smaller $\kappa$ will increase processing time, while larger $\kappa$ may make it more difficult to match fast camera motion. We found $\kappa=20$ to be a good compromise in our test sequences.

\paragraph*{3D scene points}
To recover 3D points across the scene, we solve a keypoint reconstruction problem that is independent per time step. Taking as fixed the recovered camera poses for each video frame, we match 2D keypoints between frames with the same timestamp, then reconstruct a set of sparse 3D points per time step. This is our key to handling dynamic scenes: as 2D keypoints are not matched in time, moving objects are correctly recovered in space even if their motion makes matching over time difficult. However, this knowingly produces temporal inconsistencies; we will recover from these errors in novel view synthesis where it is easier to enforce consistency (Sec.~\ref{sec:rendering}).

\paragraph*{Post processing} Finally, we increase the density of our point matches using PatchMatch~\cite{Barnes09}, as proposed in the OpenMVS\footnote{\url{https://github.com/cdcseacave/openMVS}} and COLMAP\footnote{\url{https://colmap.github.io/}}~\cite{schonberger2016structure} frameworks. This process splats points to each view and assigns colors to the 3-D point cloud.

\subsection{Novel depth and novel view rendering}
\label{sec:rendering}

Our SfM recovers a set of camera poses and an RGB 3D point cloud per time step. However, at this stage of our algorithm, projecting these points to a novel view still leaves large regions of empty space. To synthesize more realistic views, we diffuse these points in depth and RGB in the new view in image space while enforcing spatio-temporal constraints.

\paragraph*{Notation} We will often warp the content of a frame $I_*$ into the domain of the novel view $I_t$: this reprojection is computed using the extrinsic and intrinsic parameters of both reprojected frames and virtual camera, as well as the depth map $D$ with values $d$ associated to each pixel. 
We will denote it  $I_*^{proj}(x)=I_*(\mathbf{C}_*\mathbf{C}^{-1}_t(x_t,d_t))$, where $\mathbf{C}^{-1}(x,d)$ is the image plane to world coordinate system transformation of the pixel location $x$ given its depth value $d$. 
We also denote by $\hat \cdot$ a sparse map. The sparse depth map obtained by projecting the sparse point cloud into frame $t$ of the new virtual camera path is then $\hat D_{t}$ and its corresponding sparse color image $\hat I_t$.

\paragraph*{Algorithm progression} We wish to warp a frame $I_{s,t}$ to the novel view $I_t$  to be blended into a final novel view. For this, we need both the estimated camera poses and the dense depth maps $D_t$, which are yet to be computed. But, to properly constrain the diffusion of the sparse depth values $\hat D_t$, recovered in Sec.~\ref{sec:pose_estimate}, we need RGB information from the virtual camera's point of view. Thus, we jointly solve for the depth maps $D_t$ and color images $I_t$ by minimizing the energy functional:
\begin{equation}
\label{eq:functional}
    E = E_D + E_I.
\end{equation}
The functional relates terms constraining the depth map ($E_D$) to terms constraining the color image ($E_I$) by weights that guide the diffusion process. We solve $E$ iteratively: we first solve for the depth map $D_t$ while fixing the color values $I_t$, and then conversely we fix the depth values and solve for color. This avoids having to solve a nonlinear system of equations, and lets us use slightly-improved depth values to warp the input frames at each step. This betters the estimate of the rendered RGB image, which in turn constrains the diffusion of the depth.


\input{figures/method_depthmap}
\begin{figure*}
    \centering
    
    \begin{tabular}{c@{\hspace{1mm}}c@{\hspace{1mm}}c@{\hspace{1mm}}c}
        \includegraphics[width=0.24\linewidth]{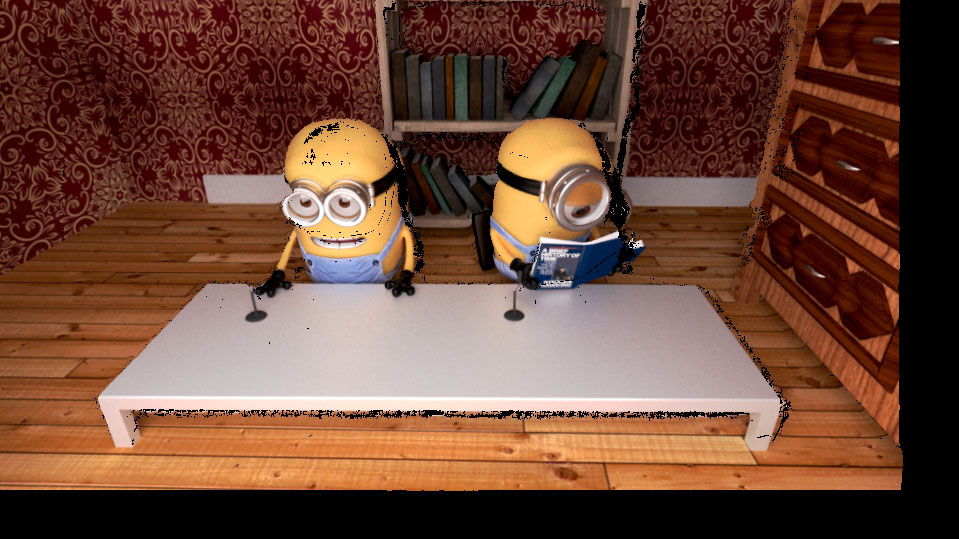} &
        \includegraphics[width=0.24\linewidth]{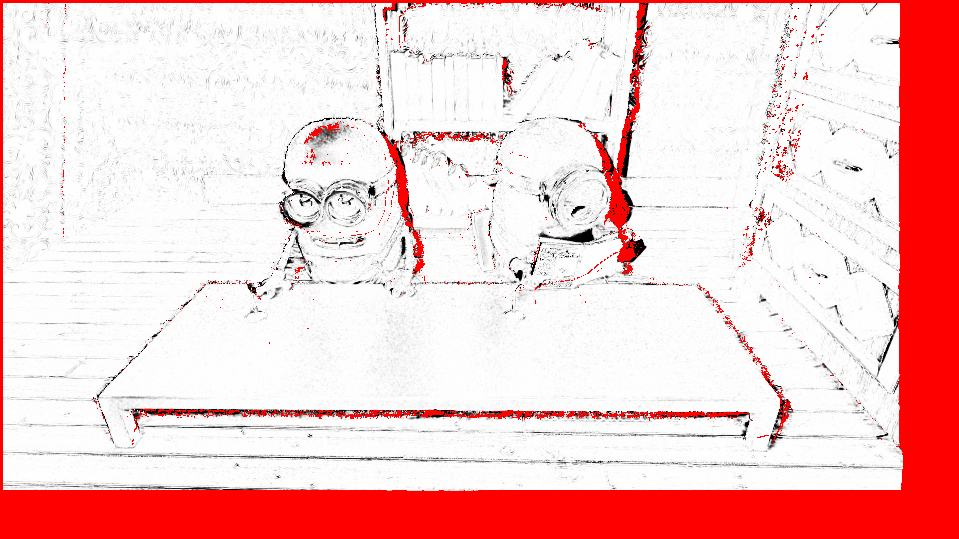} &
        \includegraphics[width=0.24\linewidth]{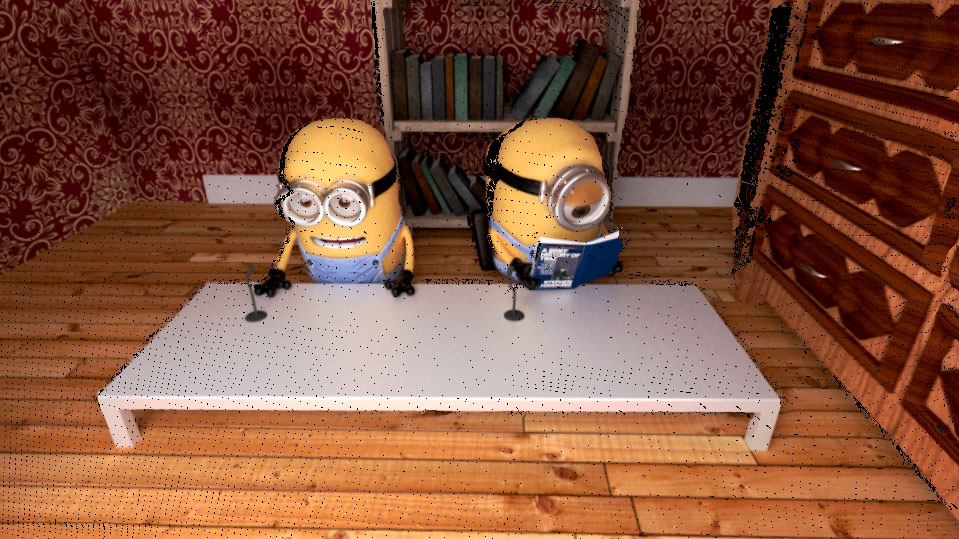} &
        \includegraphics[width=0.24\linewidth]{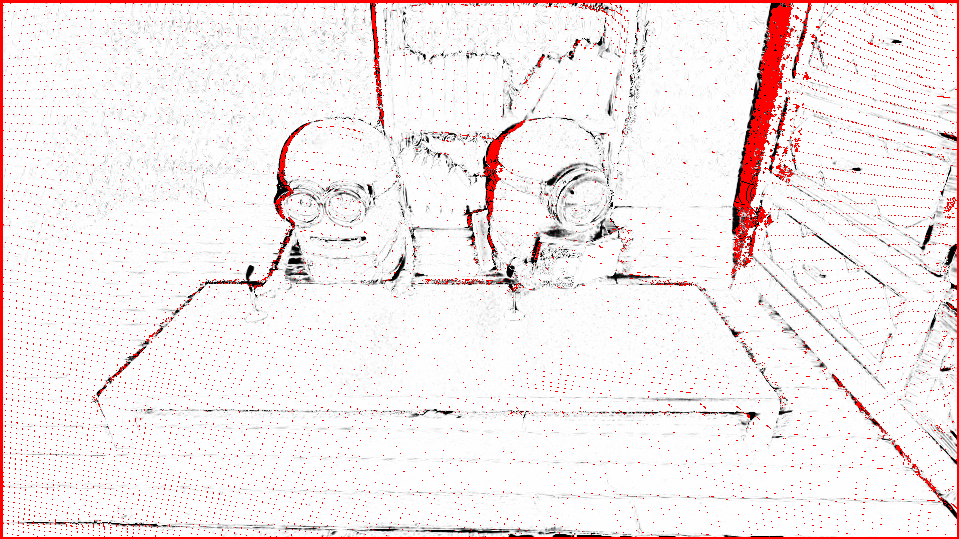} \\
        \includegraphics[width=0.24\linewidth]{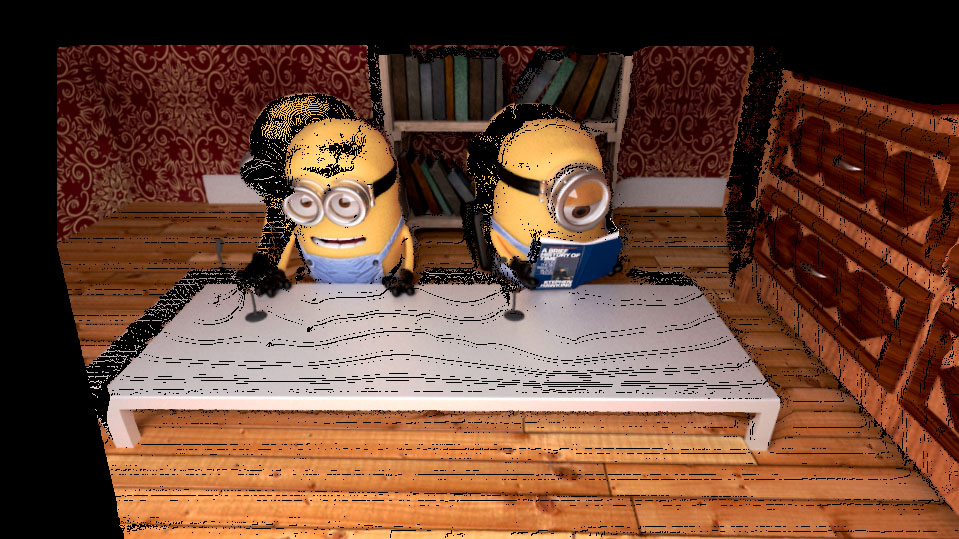} &
        \includegraphics[width=0.24\linewidth]{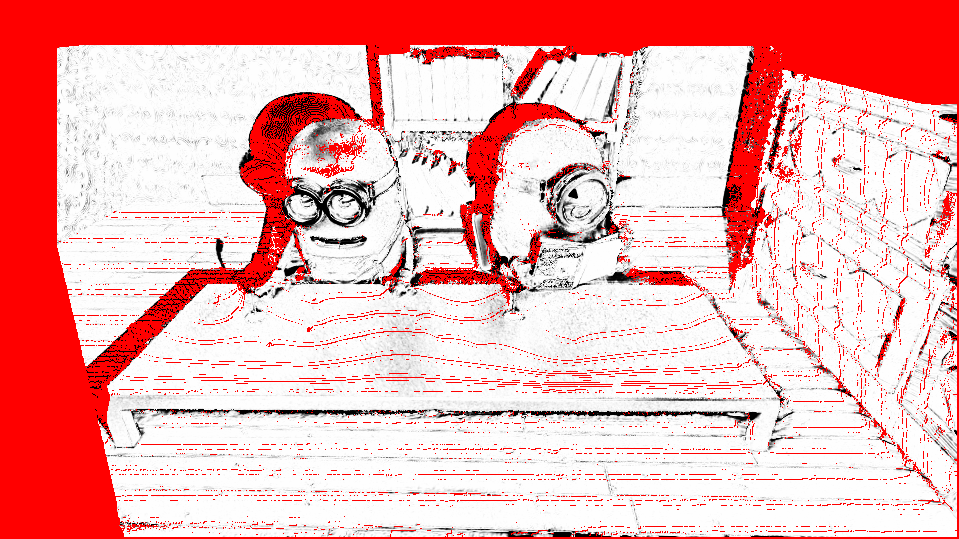} &
        \includegraphics[width=0.24\linewidth]{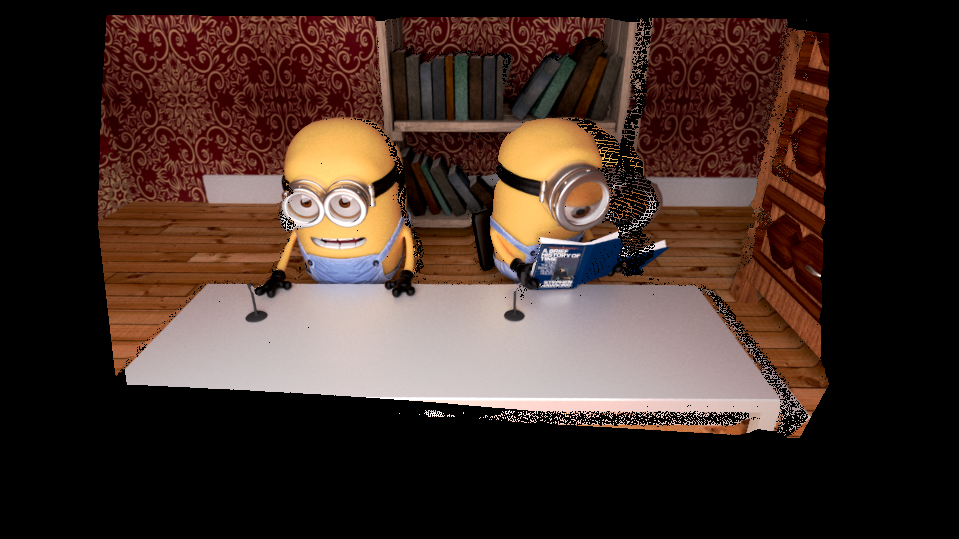} &
        \includegraphics[width=0.24\linewidth]{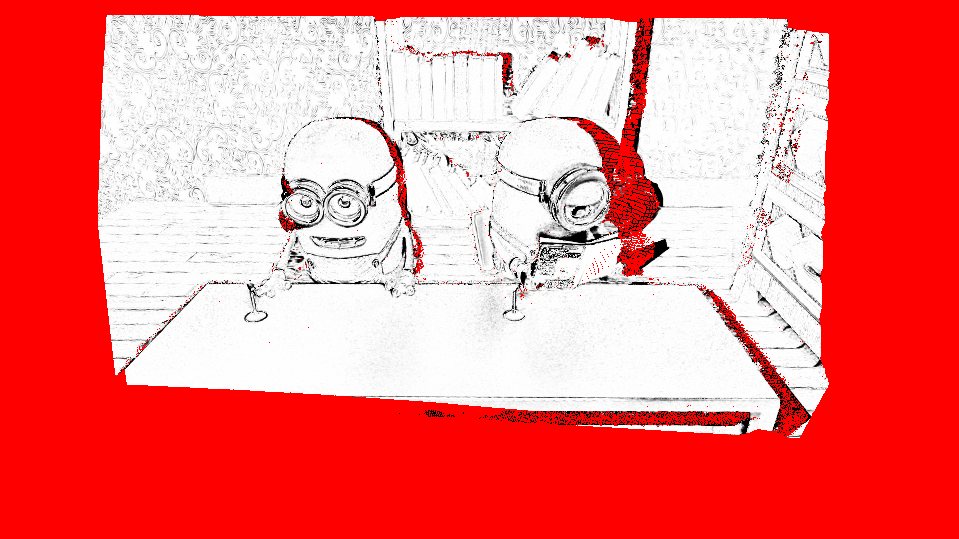} \\
    \end{tabular}
    
    \caption{Four closest input images $I_{s,t}$ projected onto the virtual camera's view point alongside their corresponding weight maps $w_P$ (Eq.~\ref{eq:wP}).}
    \label{fig:reprojected}
\end{figure*}
\input{figures/method_temp_cons}

\paragraph*{Depth diffusion} We project the sparse point cloud into the novel view, creating the sparse depth map $\hat D_{t}$ as an initialization. Then, we densify it by minimizing the following energy:
\begin{align}
    \begin{split}
        \label{eq:rendering_depth}
        E_D = & \int_{x\in {\Omega}}w_D(x,t)\|\nabla D_t (x)\|^2 \text{d}x \\  + \lambda_{PC} & \int_{x\in {\Omega}}w_{\hat D}(x,t)\ \|D_t(x) - \hat D_{t}(x)\|^2\text{d}x.  
    \end{split}
\end{align}

The first integral is a smoothness term controlled by weight $w_D$. 
We wish diffusion to decrease around color edges to produce sharp results.
We also wish diffusion of depth values to increase when the colors from reprojected input views are similar.
As such, we define $w_D$ as:
\begin{equation}
\label{eq:wD}  
    w_D(x,t) = \frac{1}{\|\nabla I_t\|^2\sum_{s = 1}^{n}\sigma_{vis}^s(x,t)} \sum_{s = 1}^n w_P^s(x,t),
\end{equation}
where $\nicefrac{1}{\|\nabla I_t\|^2}$ modulates depth diffusion around color edges, and $\nicefrac{1}{\sum_{s = 1}^{n}\sigma_{vis}^s(x,t)}$ is a normalization factor that accounts for each pixel's visibility in the novel view. As both the visibility term $\sigma_{vis}$ and the projection weight $w_P^s$ pertain more to the color diffusion process, we will defined them later on in Eq.~\ref{eq:wP}.

The second integral reduces the weight of sparse 3D points that are occluded from the point of view of the virtual camera or are erroneously reconstructed. For this, we relax the constraint of $D_t$ where it exactly matches the projected sparse point cloud:
\begin{equation}
    \label{eq:wH}
    w_{\hat D}(x,t) = \exp\left(-\frac{\|\hat I_{t}(x) - I_{t}(x)\|^2}{2\sigma^2}\right).
\end{equation}

In Figure~\ref{fig:depthmap}, we show example weight maps $w_{\hat D}$ and $w_D$ that govern the depth diffusion process, as defined in Eqs.~\ref{eq:wD} and \ref{eq:wH}.

There are three parameters in this diffusion process: $\sigma$ controls the soft occlusion tolerance, and we set $\sigma=0.075$ in all our experiments; the sparse point cloud attachment weight $\lambda_{PC}$, which we set in the range $\lambda_{PC} = 0.25\text{--}2$; and the temporal consistency term set in the range $\lambda_T = 0.01\text{--}0.1$.

\paragraph*{Color diffusion} Given depth map $D_t$, we initialize the RGB image to a projection of the color in the input point cloud. Then, we densify it by minimizing the following diffusion energy:
\begin{align}
    \begin{split}
    \label{eq:rendering_color}
        E_I = & \int_{x\in {\Omega}}\|\nabla I_t\|^2\\ 
        +\sum_{s=1}^n & \int_{x\in {\Omega}}\lambda_{P}w_P^s(x,t)\|I_t(x) - I_{s,t}^{proj}(x)\|^2 \text{d}x\\
        +\sum_{s=1}^n & \int_{x\in {\Omega}}\lambda_Gw_P^s(x,t)\|\nabla I_t(x) - \nabla I_{s,t}^{proj}(x)\|^2 \text{d}x
    \end{split}
\end{align}

The first integral encourages smooth gradients over the intensity of the novel view, which aids blending of the projected input images especially along their borders. 
The second integral constrains the RGB intensities and $I_t$ to be close to the intensities of $I_{s,t}^{proj}$, 
and the third integral constrains the RGB gradients similarly.
They are both modulated by the weight
\begin{equation}
    \label{eq:wP}
    w_P^s(x,t) = \sigma_{vis}^s(x,t)\exp\left(-\frac{\|I_{s,t}^{proj}(x) - I_t(x)\|^2}{2\sigma^2}\right),
\end{equation}
which measures the agreement of each warped input frame with the novel view. Figure \ref{fig:reprojected} shows a set of warped input frames along with their weight maps $w_P^s$.

\begin{wrapfigure}{l}{0.45\linewidth}
    \includegraphics[width=\linewidth]{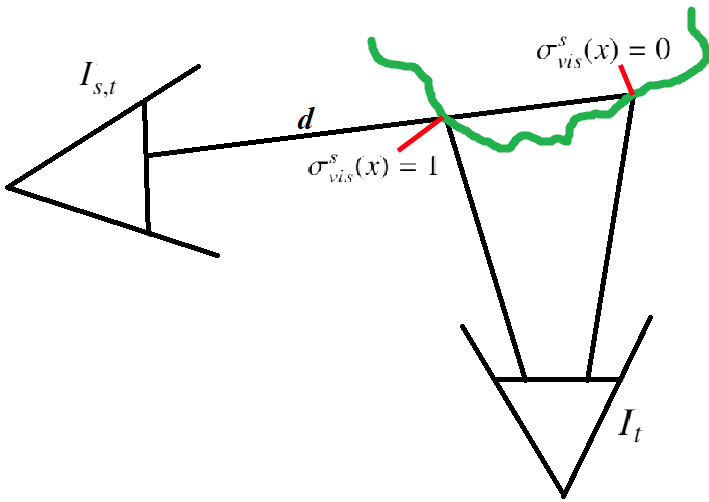}
\end{wrapfigure}
$w_P$ incorporates visibility term $\sigma_{vis}^s(x,t)$ that is $1$ for a given pixel $x$ of the novel view $I_t$ only if, out of every pixel that is projected to the same pixel location in an input image $I_{s,t}$, $x$ has the smallest depth value $d$ in the input image's coordinate frame.

The color diffusion relies on two new parameters: $\lambda_P$ and $\lambda_G$ balance the weight over the data and the gradient equality constraints. We set them both in the range $5\text{--}20$. $\sigma$ and $\lambda_T$ serve the same function and values as in the depth map diffusion.

\subsection{Temporal consistency}

We enforce temporal consistency within novel views by additional terms in $E_D$ and $E_I$. With slight abuse of notation:

\begin{align}
    \begin{split}
        \label{eq:depth_temporalconsistency}
        E_D &= \dots + \lambda_{T}\int_{x\in \Omega}w_{T}(x,t)\ \| D_t(x) - D_{t-1}^{proj}(x) \|^2 \text{d}x,
    \end{split}
\end{align}

\begin{align}
    \begin{split}
    \label{eq:color_temporalconsistency}
        E_I &= \dots + \lambda_{T}\int_{x\in \Omega}w_{T}(x,t)\ \| I_t(x) - I_{t-1}^{proj}(x)\|^2 \text{d}x.
    \end{split}
\end{align}

These terms constrain depth $D_t$ to remain similar to the warped previous depth $d_{t-1}$, and for color $I_t$ similarly.
This constraint is relaxed by a weight
\begin{equation}
\label{eq:wT}
    w_T(x,t) = \frac{1}{n}\sum_{s=1}^{n}\exp\left(-\frac{\|I_{t-1}^{proj}(x) - I_{s,t}^{proj}(x)\|^2}{2\sigma^2}\right)
\end{equation}
for pixels for which an agreement in color was not reached. This is expected in regions containing motion because the depth values of frame $t-1$ may be invalid, as is the case around the mouth of the character on the left in Figure~\ref{fig:temp_cons}. $w_T$ allows the computation of depth and color values of these pixels to rely more freely on the other terms of the functional, like the data term of the depth or the color of the warped input images.

\subsection{Implementation details}

To avoid using input frames that are far away from the novel camera's view, we rank each input camera based on its distance from the novel camera according to the following formula:
\begin{equation}
    \label{eq:rF}
	\!r_F(s)\!=\!\frac{1}{\|C_t-C_{s,t}\|^2}\!\exp\!\!\left(-\frac{\arccos{((\text{tr}(R_tR_{s,t}^T)-1)/2)}}{2\pi\sigma^2}\right)
\end{equation}
This penalizes frames that are either far in center or in viewing direction from the novel view.
Then, we use the first $n=4$ ranked input frames to minimize the functional Eq.~ \ref{eq:functional}.

For efficiency, we also proceed in a multiscale fashion: we solve for depth and color at a coarse resolution, and then use these to initialize a finer resolution---our lowest level is 1/64 of the original frame size. Finally, we also proceed in a streaming manner: we reproject the previous frame's depth and color (denoted as $D_{t-1}^{proj}$ and $I_{t-1}^{proj}$) into the current virtual camera pose for use within the temporal consistency constraint.
\section{Experiments and Results}

\subsection{Dataset Sequences}
\paragraph*{Real-world existing dataset}
We exploit existing datasets used in the context of novel view synthesis, all of them captured using camera arrays:
\begin{itemize}[itemsep=0.5pt]
    \item Jumping \cite{yoon20}: A group of four people jump (12 cameras).
    \item Skating \cite{yoon20}: A person rides a skateboard (12 cameras).
    \item Playground \cite{yoon20}: A person flies a dinosaur balloon (12 cameras). 
    \item Umbrella \cite{yoon20}: A person opens and rotates an umbrella (12 cameras).
    \item DynamicFace \cite{yoon20}: A person of making faces (12 cameras).
    \item Breakdancers \cite{Zitnick2004}: A person break dancing in front of 4 people (8 cameras).
\end{itemize}

\paragraph*{Custom dataset}
We test our algorithm on three 100-frame real world sequences that we acquired each with five cameras at 1920$\times$1080 resolution. The cameras were hand held or set on tripods (Canon Rebel EOS T7i). 
We additionally generate a synthetic sequence using 11 input cameras to compare to ground truth RGB and depth estimation from a 12th camera. 
Our sequences are:
\begin{itemize}[itemsep=0.5pt]
    \item Cat and dog: Two pet animatronics,
    \item Minions (synthetic): a rendering of two characters laughing behind a table, 
    \item Elephant wiggle: A puppet hanging by a wire, and
    \item Drone: A drone hanging by a wire.
\end{itemize}

\begin{figure*}
    \centering
    
    \newcommand{\imgw}{0.20\linewidth}
    
    \begin{tabular}{@{}c@{\hspace{1mm}}c@{\hspace{1mm}}c@{\hspace{1mm}}c@{}}
        \includegraphics[width=\imgw]{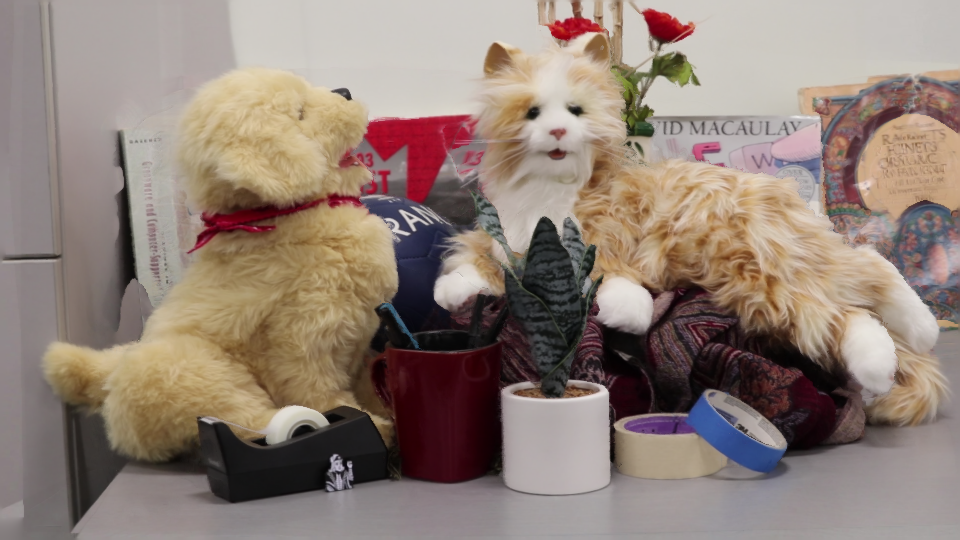}&
        \includegraphics[width=\imgw]{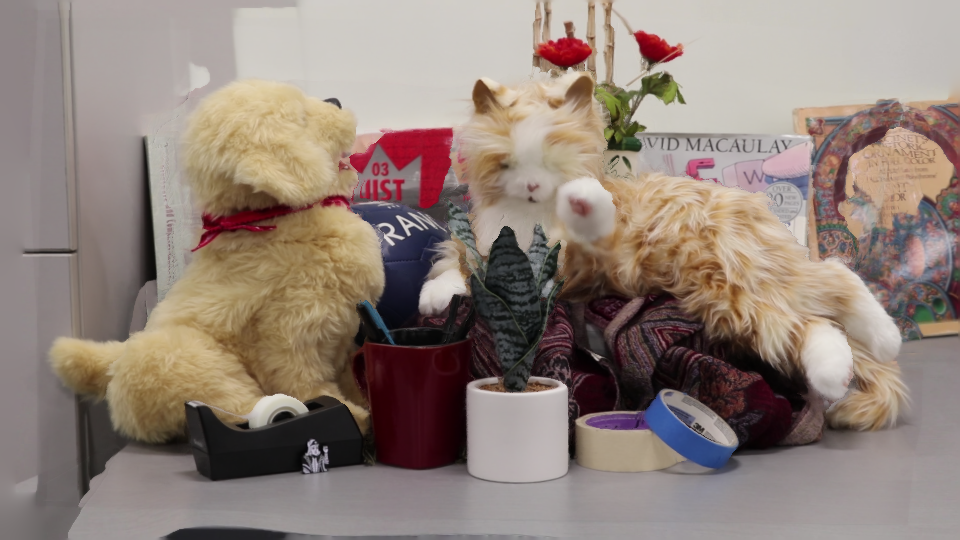}&
        \includegraphics[width=\imgw]{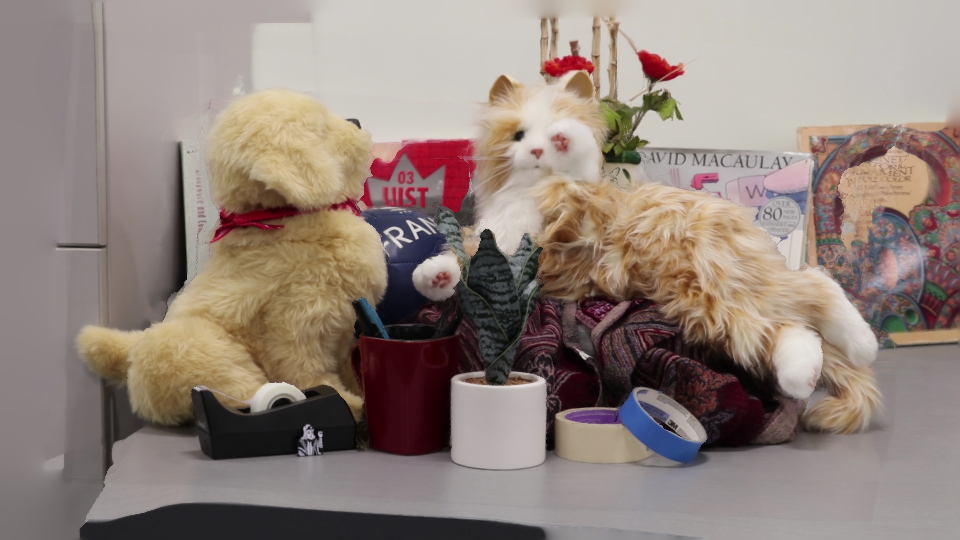}&
        \includegraphics[width=\imgw]{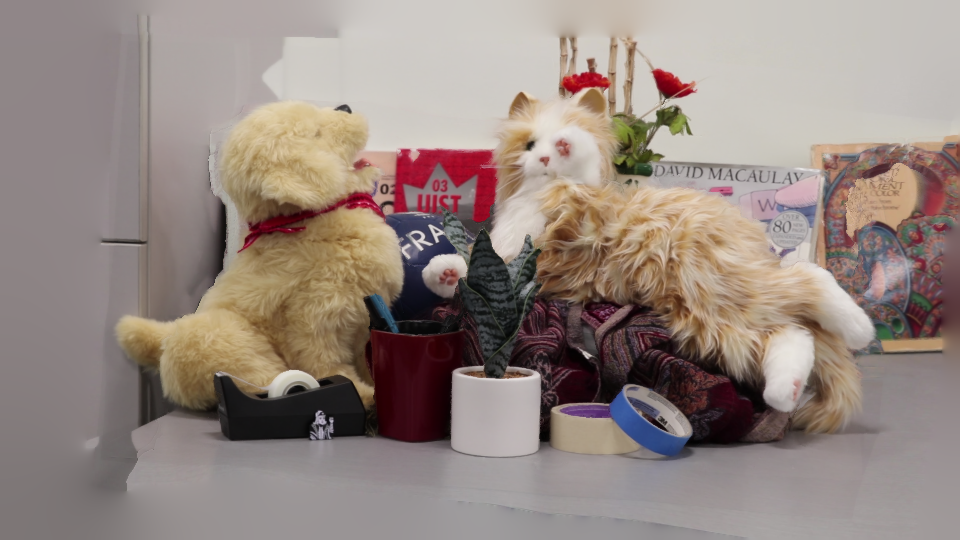}\\
        
        \includegraphics[width=\imgw]{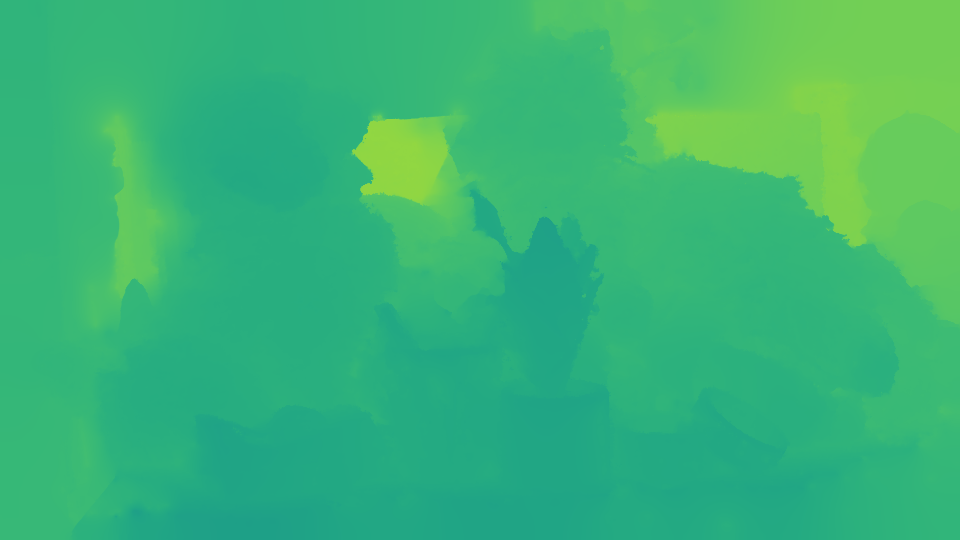}&
        \includegraphics[width=\imgw]{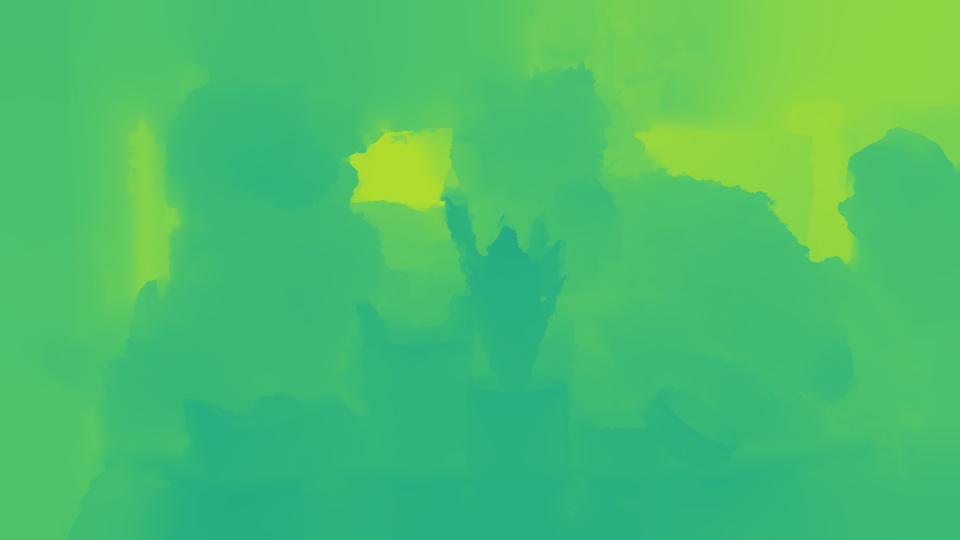}&
        \includegraphics[width=\imgw]{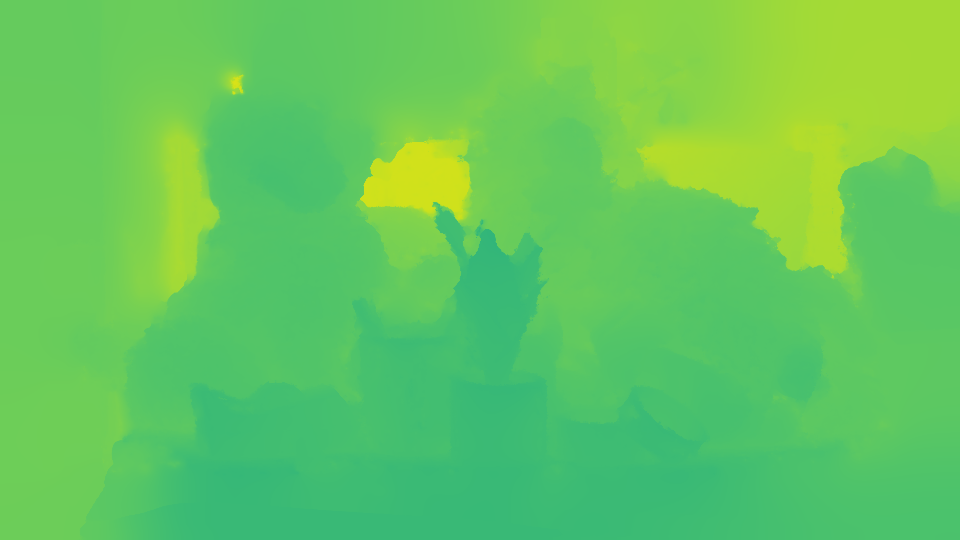}&
        \includegraphics[width=\imgw]{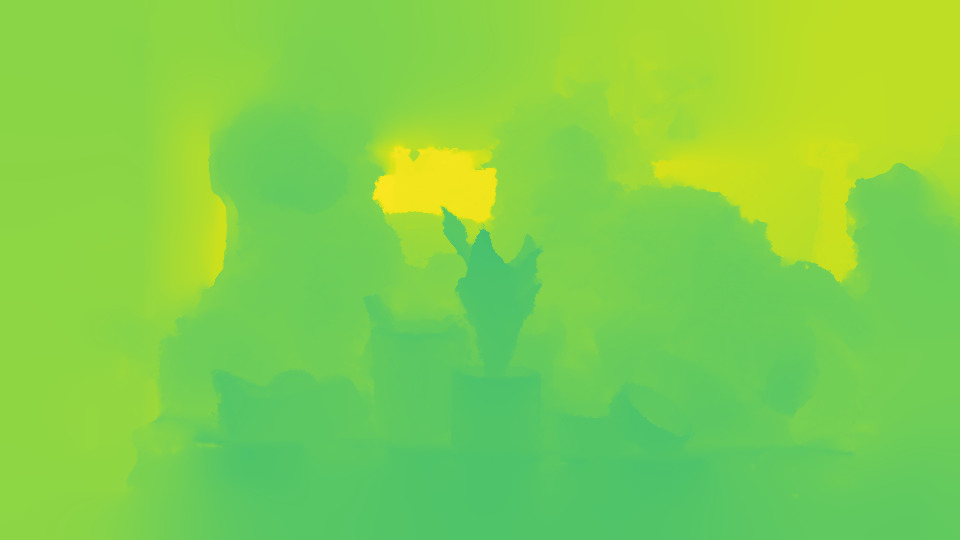}\\

        \includegraphics[width=\imgw]{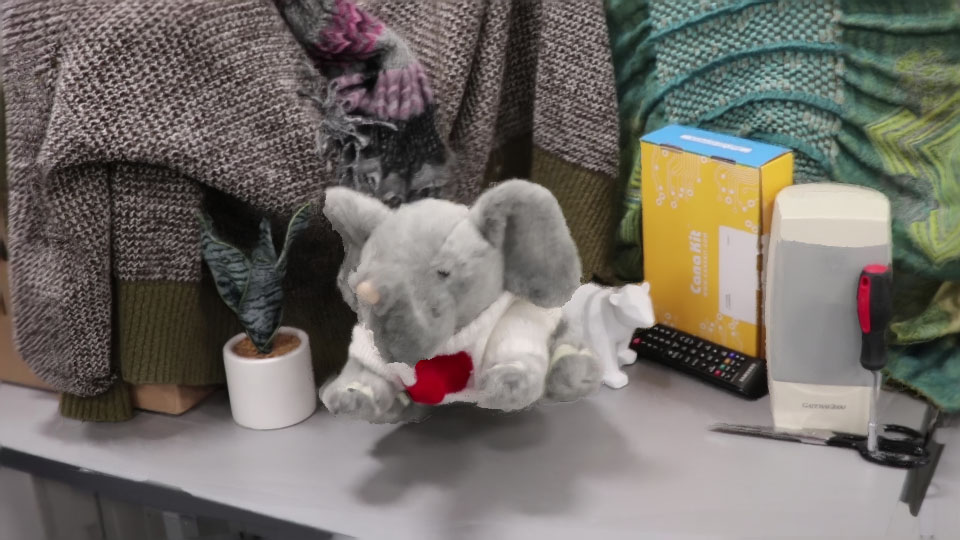}&
        \includegraphics[width=\imgw]{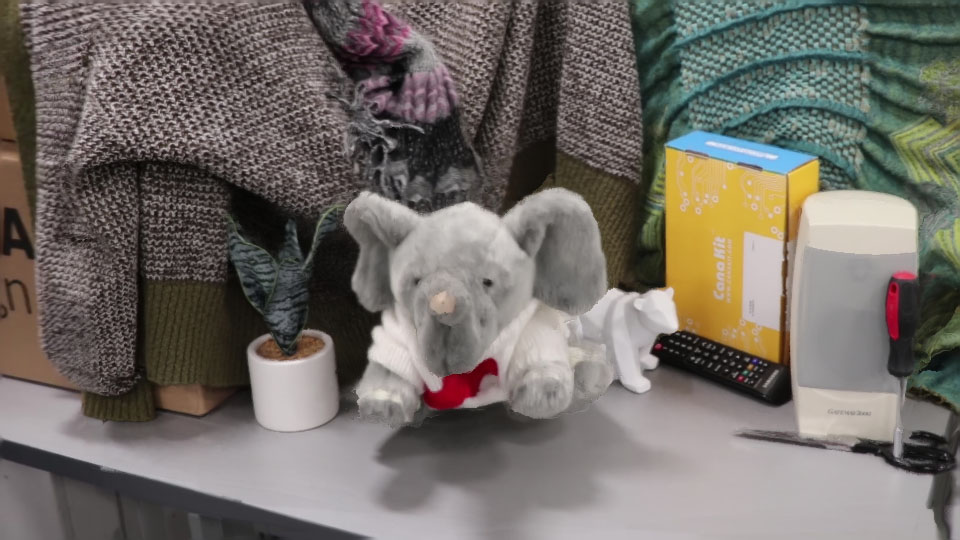}&
        \includegraphics[width=\imgw]{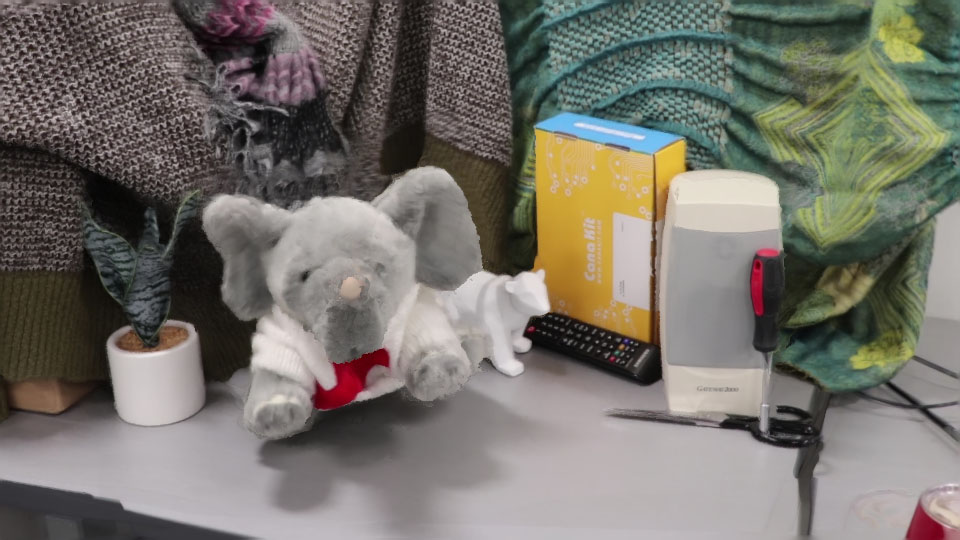}&
        \includegraphics[width=\imgw]{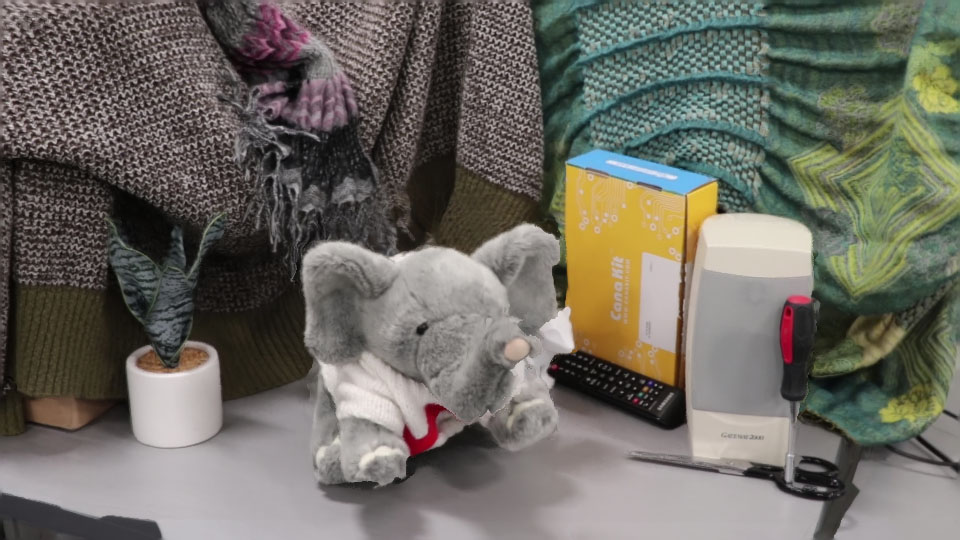}\\
        
        \includegraphics[width=\imgw]{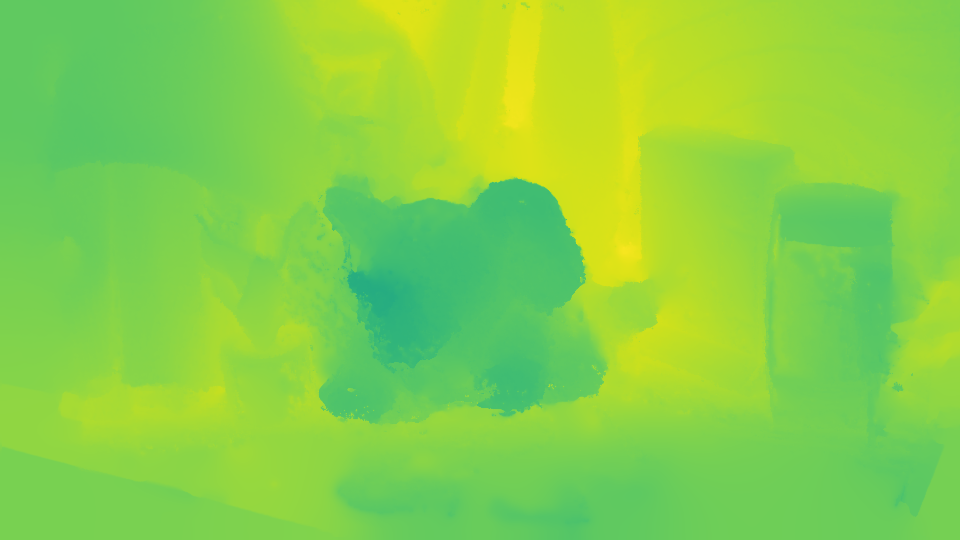}&
        \includegraphics[width=\imgw]{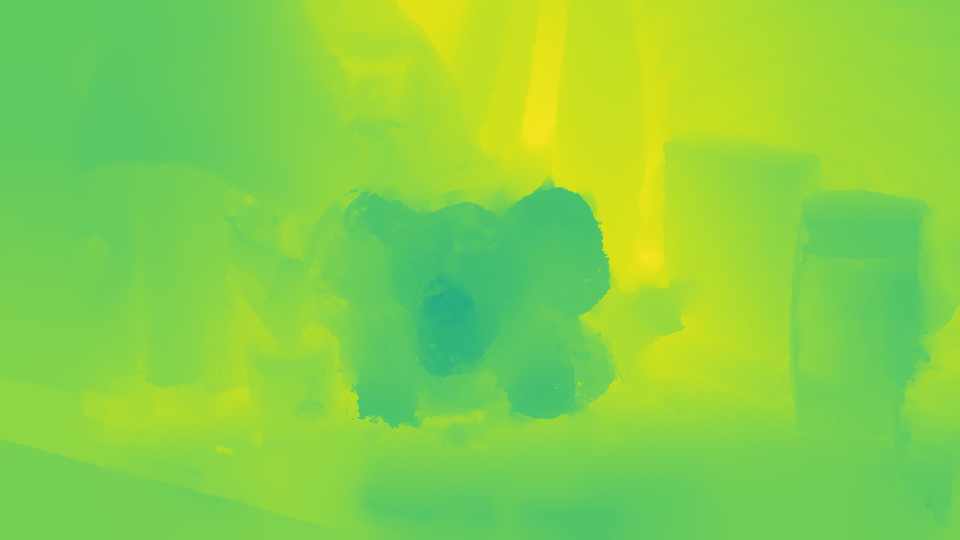}&
        \includegraphics[width=\imgw]{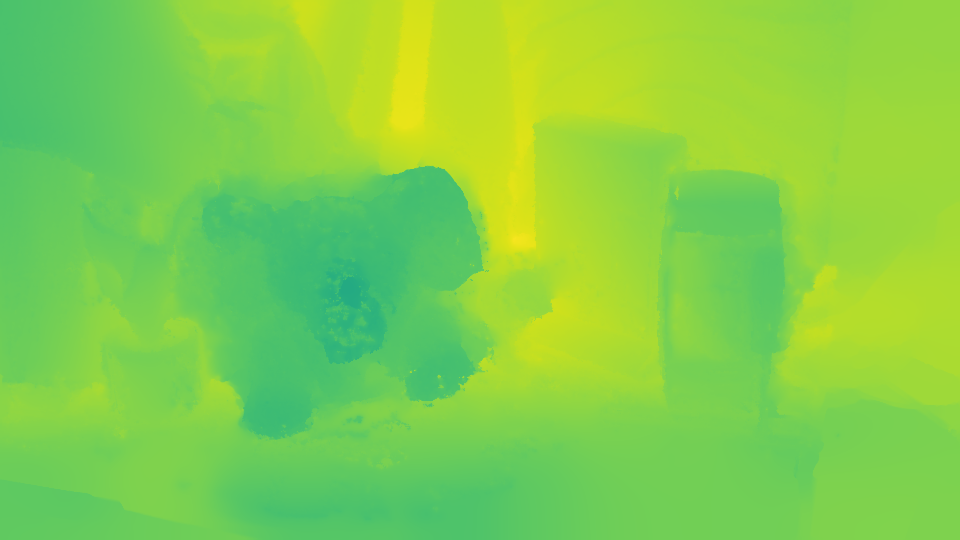}&
        \includegraphics[width=\imgw]{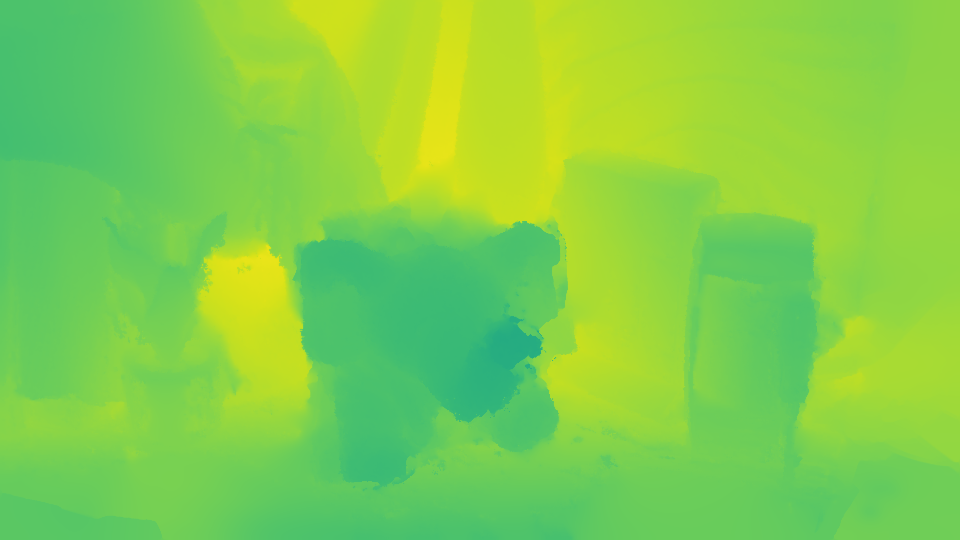}\\
    \end{tabular}
    \caption{Color and depth results for Cat\&dog and Elephant-wiggle scenes.}
    \label{fig:exp_results}
\end{figure*}

Figure \ref{fig:exp_results} shows rendered frames from novel views and corresponding depth maps for the Cat and dog, and the Elephant wiggle sequences. While some artifacts remain in the depth video, the generation of the final novel view RGB rendered sequence is robust to these and has fewer artifacts. Note that the borders of the view partially appear blurry when there is insufficient field of view overlap between input videos.

\subsection{Ablation study}
\input{figures/exp_ablation_2rows}
We ablate our SfM method using the synthetic dataset with moving objects, where points are known to be either static or dynamic (Table~\ref{tab:exp_sfm_ablation}). We compared the recovered pose over 30 timestamps and 11 cameras. First, we compare against a naive SfM approach that solves for all frames simultaneously without consideration of dynamic objects. Next, some methods rely on segmenting out moving objects to cope with dynamic scenes~\cite{Mustafa16,Mustafa19}. To compare to this idea, we created a segmentation-based SfM baseline from the naive SfM by performing reconstruction only from points that are known to be static using perfect ground truth masks.
While the segmentation slightly aids the recovery of camera positions, its positive effect is not clear on the 3D reconstruction, even though the dynamic object segmentation is a pixel accurate ground truth.
Against both baselines, our method can make better use of dynamic points to more accurately recover scene points.
Finally, we compare against the non-smoothed camera path version of our approach. While the rotation error decreases, the positional error slightly increases. Overall, we found smoothing to provide better final results.

\subsection{Novel depth and view comparisons}
\begin{table*}[t]
    \centering
    \begin{tabular}{l | r r r p{3cm} p{2cm} r}
        \toprule
         Method & Scenes & Min. Views & Training Time & Preprocess time per frame & Render time & Figure \# \\
        \midrule
         Deep Blending \cite{deepblending} & Static & 4 & 37\ h training & 8\ h & real time & \ref{fig:exp_db}\\
         LLFF \cite{llff} & Static & 6 & ? & 10\ min & real time & \ref{fig:exp_llff}\\
         EVS \cite{evs} & Static & 2 & ? & 10\ min & 98\ sec & \ref{fig:exp_evs}\\
         MonoCam \cite{yoon20} & Dynamic & 1 & \multicolumn{3}{c}{\emph{Authors' results, no timing info}}&\ref{fig:exp_mono}\\
         VI \cite{Zitnick2004} & Dynamic & 8 &  & \emph{no timing info} & real time &\ref{fig:exp_legacy} \\
         VVC \cite{Lipski10cgf} & Dynamic & 5 & & \emph{partially manual} & real time & \ref{fig:exp_legacy} \\
         \textbf{Ours} & \textbf{Dynamic} & \textbf{4} & & \textbf{2\ min} & \textbf{6.8\ sec} & 7--11 \\
        \bottomrule
    \end{tabular}
    \caption{Comparisons regarding scene type, minimum number of input views, and speed. LLFF and EVS use pre-trained networks, so we did not re-train them. 
    }
    \label{tab:comp_table}
\end{table*}
We compare our method to four recent methods, including deep-learning-based methods requiring external training databases: Deep Blending \cite{deepblending}, Local Light Field Fusion \cite{llff}, Extreme View Synthesis \cite{evs}, and MonoCam \cite{yoon20}. Furthermore, we use the Breakdancers scene to compare to the results provided by two older methods~\cite{Zitnick2004,Lipski10cgf} that best match our intended setup. Each of these methods work with different numbers of input views and require different amounts of processing time. Some of these methods are only intended for static scenes, and so we would expect them to produce temporally inconsistent results.
Table~\ref{tab:comp_table} summarizes these properties. In this paper, we extract frames to illustrate the comparisons; please see the accompanying video to better evaluate the differences.

\textbf{Static---Deep Blending \protect\cite{deepblending}. }
\begin{figure}[t]
    \centering
    
    \begin{tabular}{@{}c@{\hspace{1mm}}c@{}}
        \includegraphics[width=0.49\linewidth,height=2.5cm]{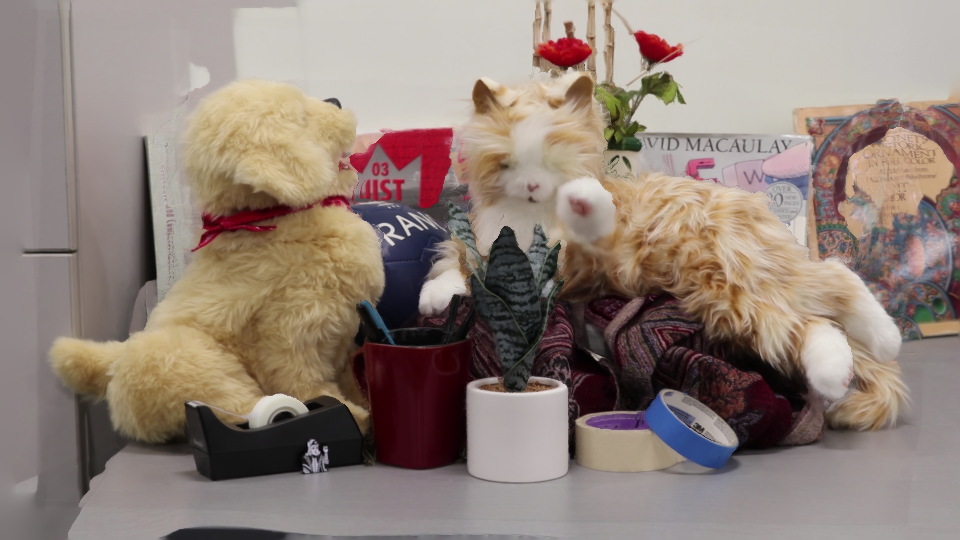}&
        \includegraphics[width=0.49\linewidth,height=2.5cm,]{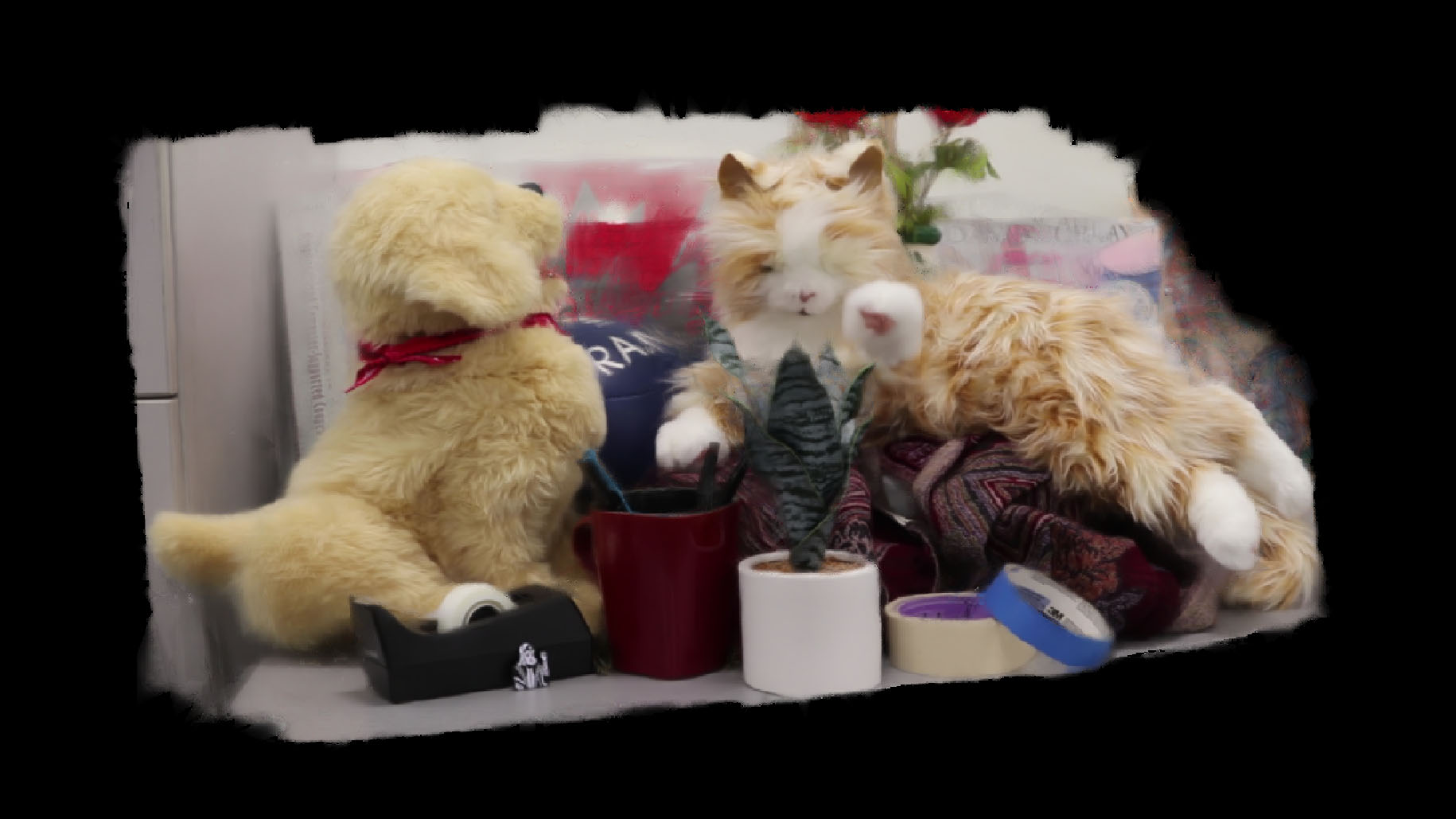}\\
        \includegraphics[width=0.49\linewidth,height=2.5cm]{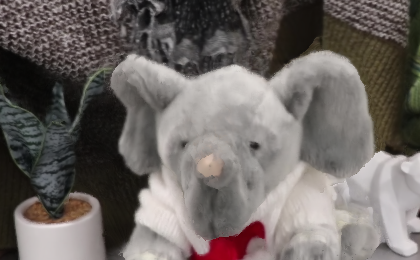}&
        \includegraphics[width=0.49\linewidth,height=2.5cm]{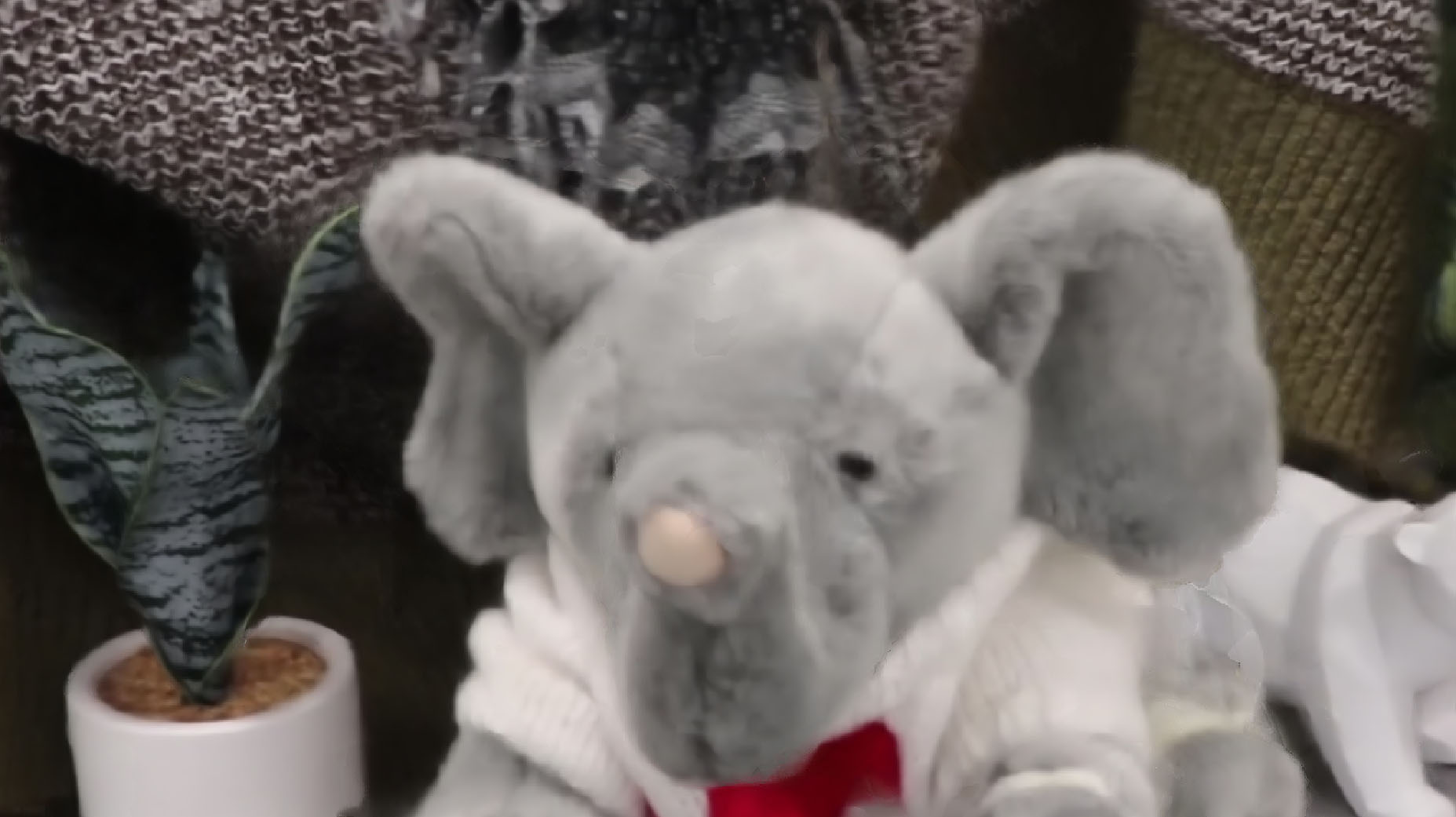}\\
        Ours & 
        DB \cite{deepblending}\\
    \end{tabular}
    
    \caption{Comparison with Deep Blending \cite{deepblending} on the Cat-Dog and Elephant-wiggle sequences.}
    \label{fig:exp_db}
\end{figure}
We compare our rendering method with Deep Blending(DB)~\cite{deepblending} which learns optimized weights for blending 4 layers of mosaic images where the first layer is composed of the best fitting pixels, the second the second best etc. based on a heuristic. For the comparison, we first reconstructed each scene separately for each time step as described in their method. Afterwards, to be able to use the same camera path as for our results, we registered each time step to our full space-time reconstruction based on the camera positions. Finally, we used the pre-trained network provided by the authors to render each frame. Figure \ref{fig:exp_db} shows DB not always being able to reconstruct marginal parts of the scenes, moreover our results appear sharper.

\textbf{Static---Extreme View Synthesis \protect\cite{evs}. }
\begin{figure*}[t]
    \centering
    
    \newcommand{\imgw}{0.16\linewidth}
    
    \begin{tabular}{@{}c@{\hspace{1mm}}c@{\hspace{1mm}}c@{\hspace{1mm}}c@{\hspace{1mm}}c@{\hspace{1mm}}c@{}}
        \includegraphics[width=\imgw]{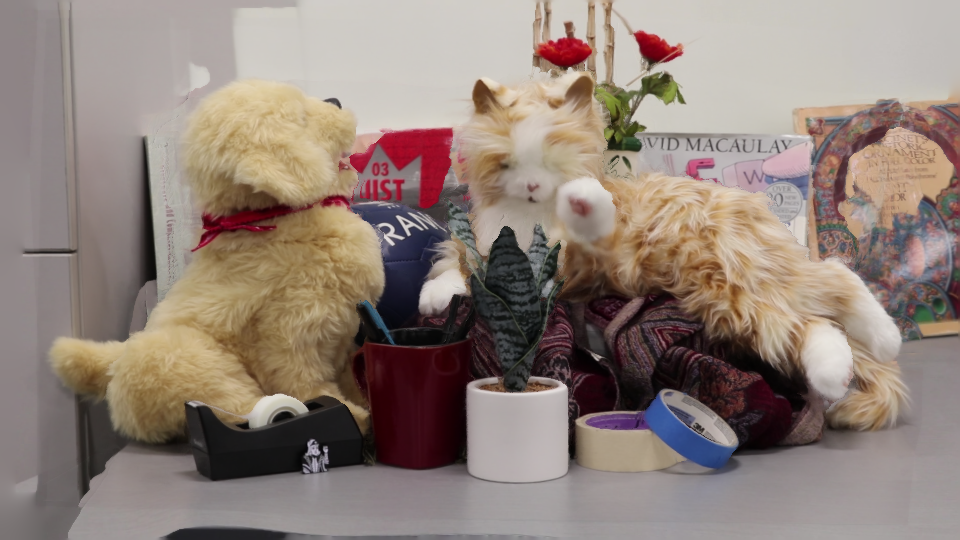}&
        \includegraphics[width=\imgw]{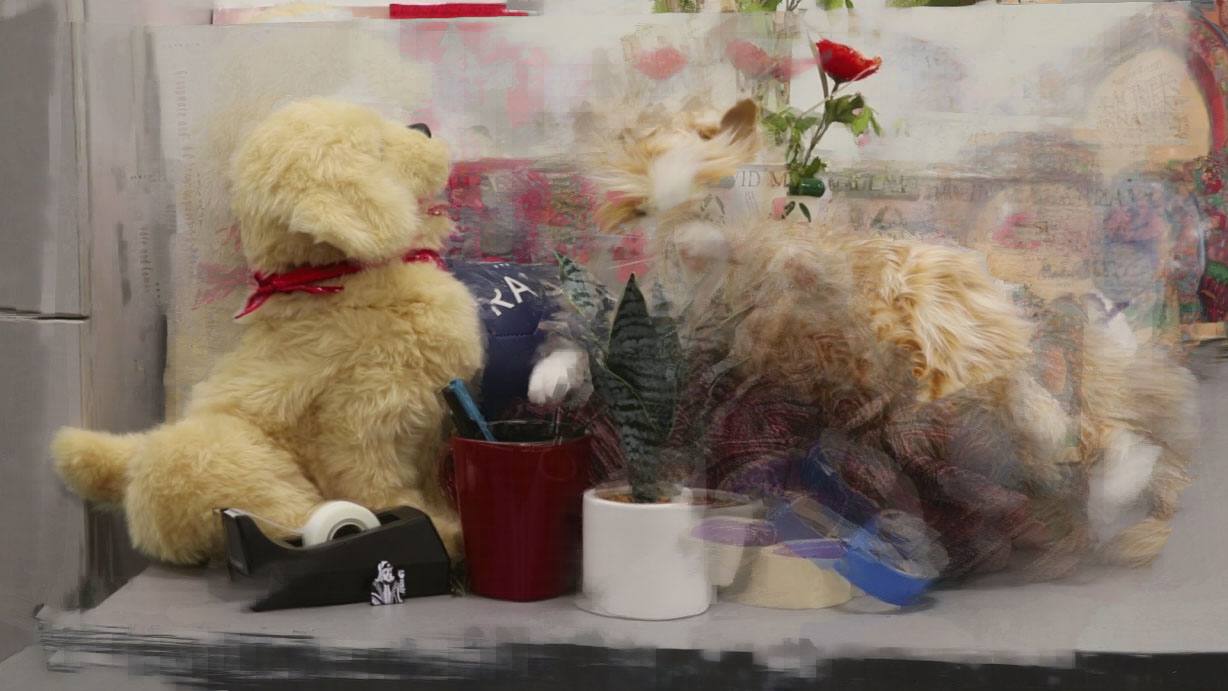}&
        \includegraphics[width=\imgw]{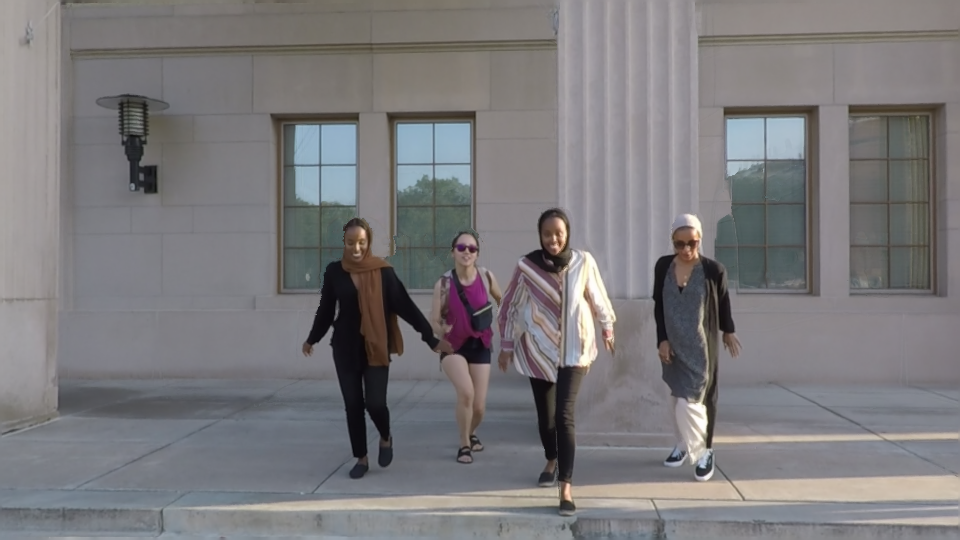}&
        \includegraphics[width=\imgw]{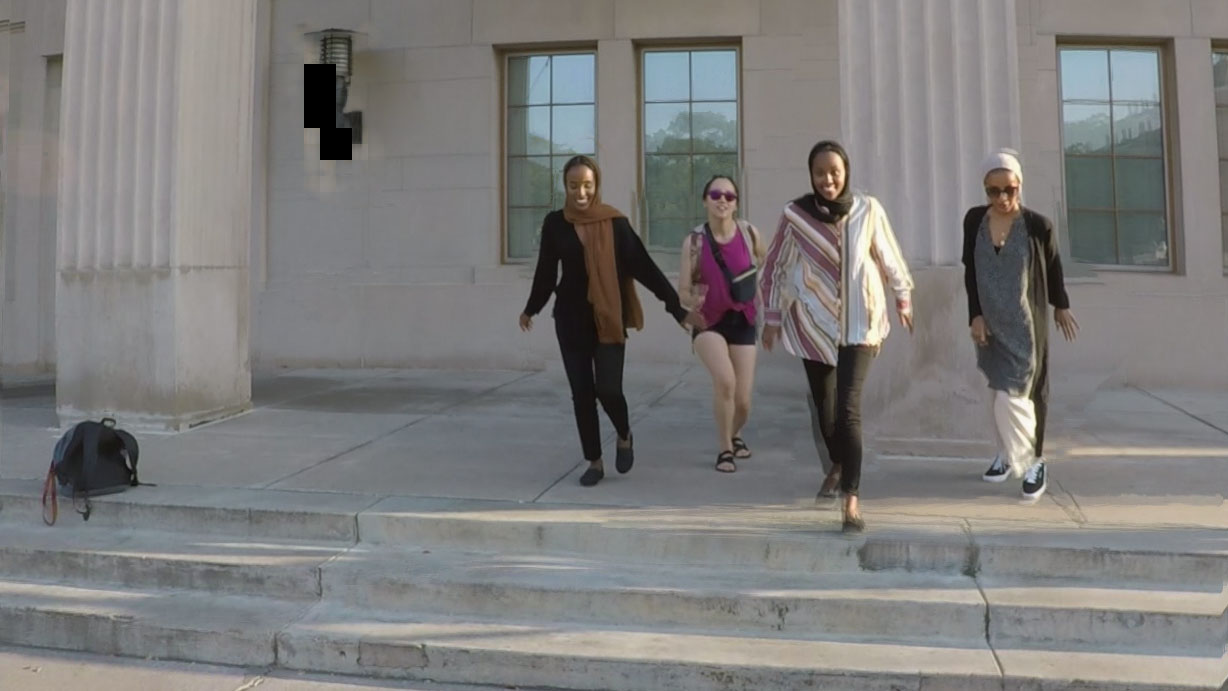}&
        \includegraphics[width=\imgw]{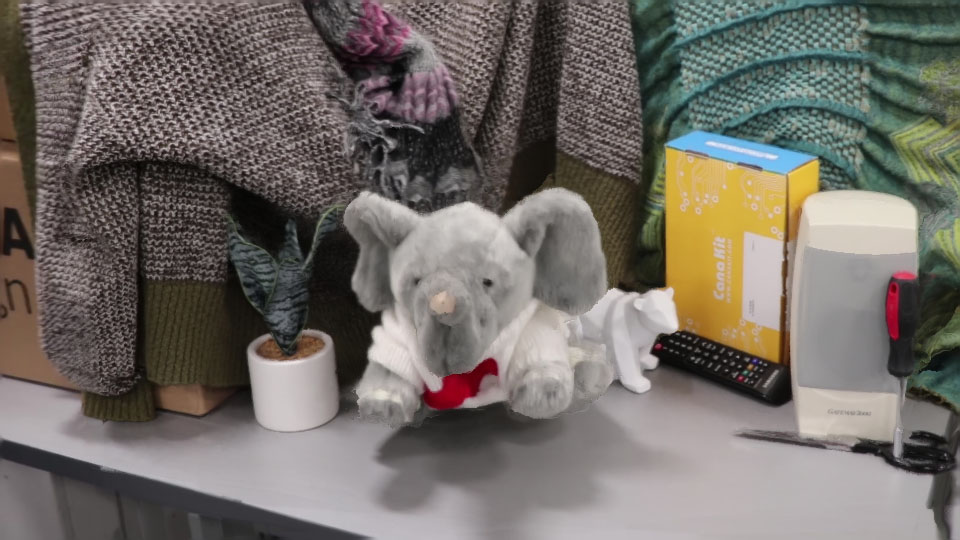}&
        \includegraphics[width=\imgw]{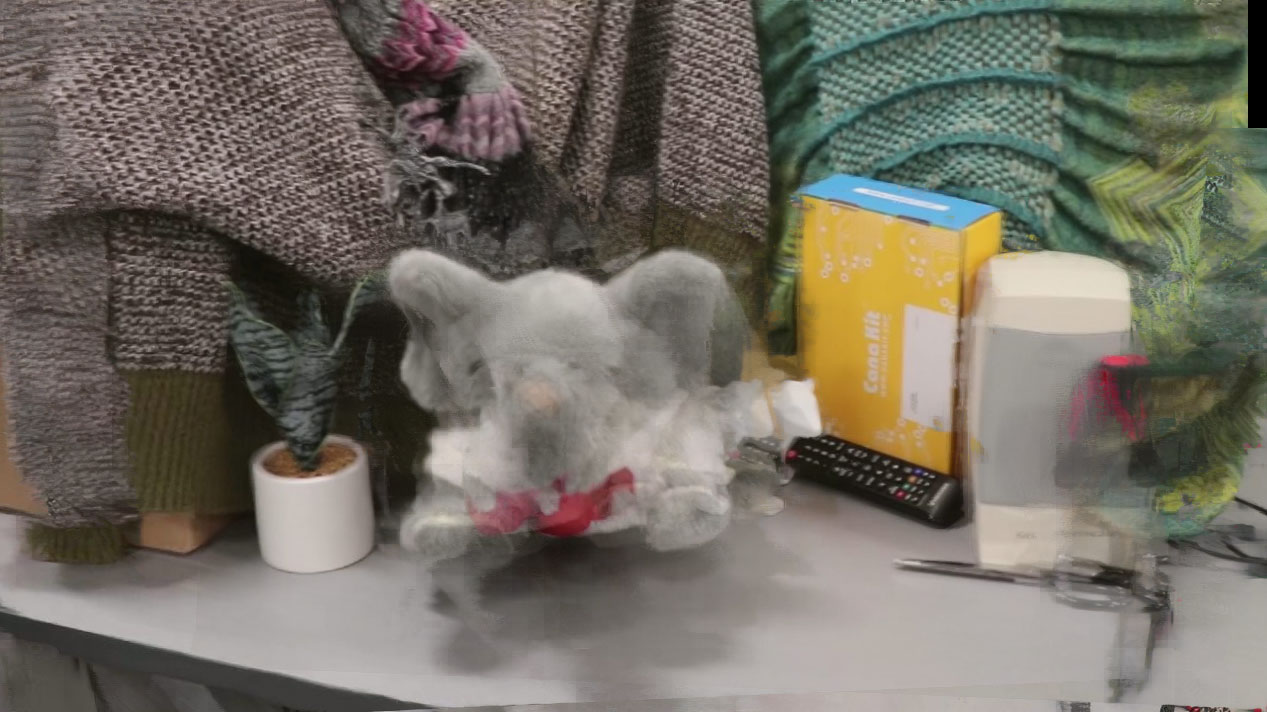} \\

        \raisebox{-.5\height}{\includegraphics[width=\imgw]{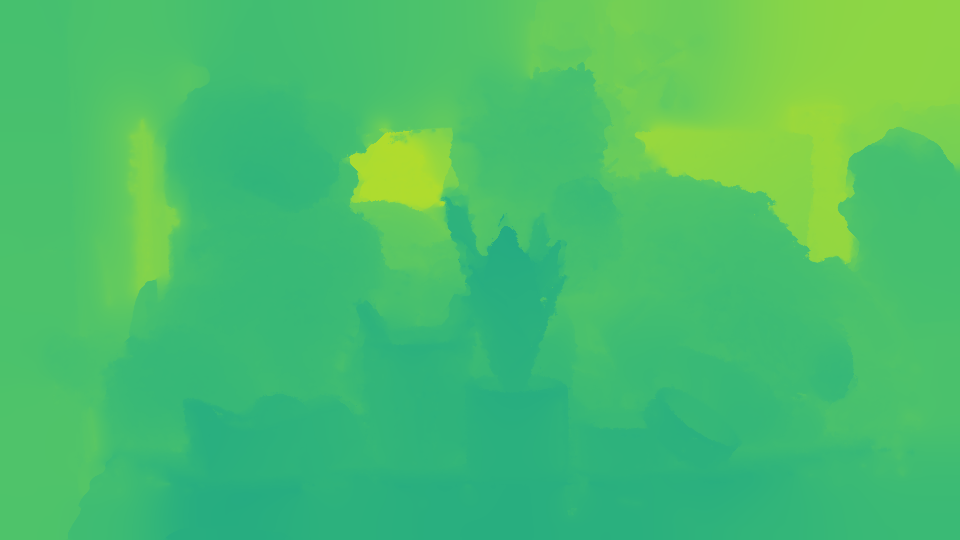}}&
        \raisebox{-.5\height}{\includegraphics[width=\imgw]{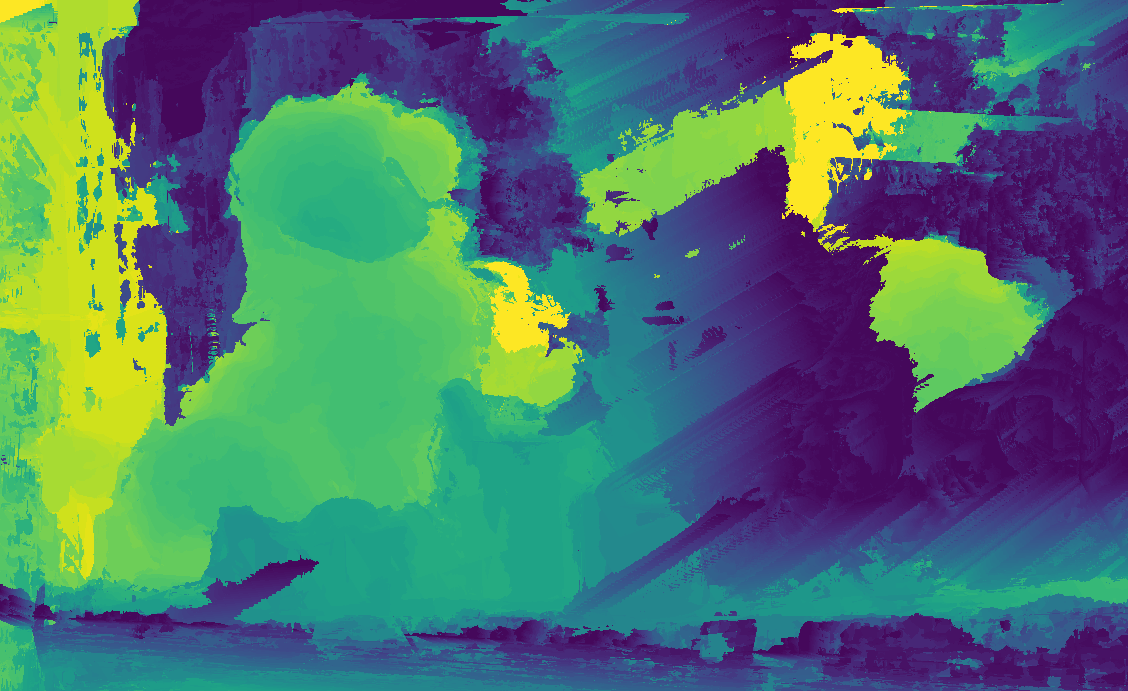}} &
        \raisebox{-.5\height}{\includegraphics[width=\imgw]{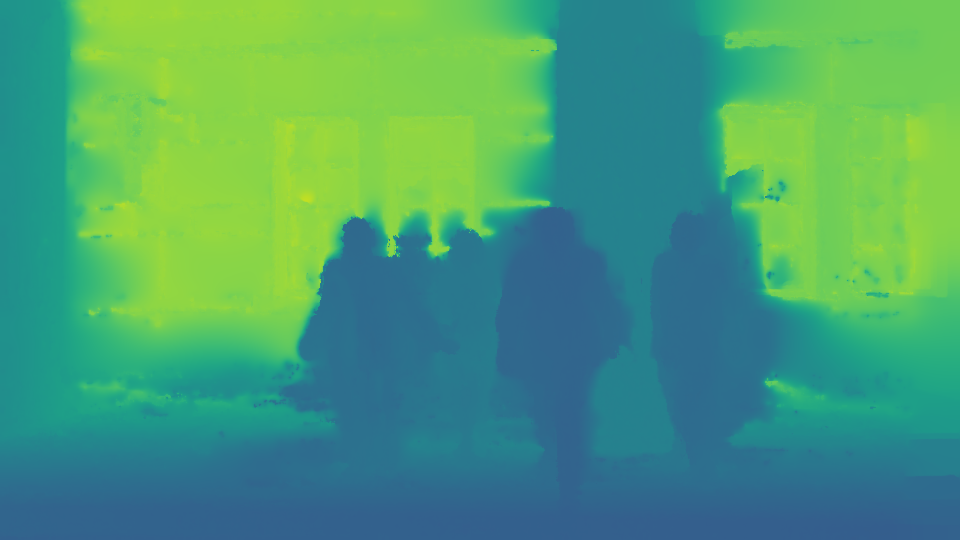}}&
        \raisebox{-.5\height}{\includegraphics[width=\imgw]{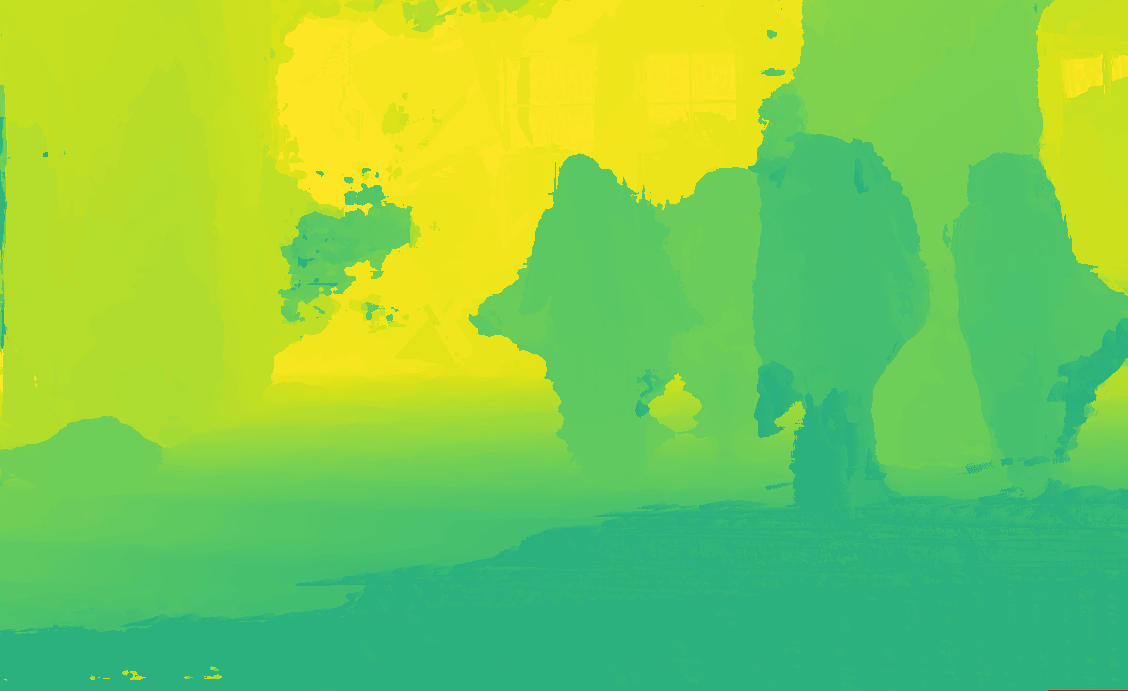}} &
        \raisebox{-.5\height}{\includegraphics[width=\imgw]{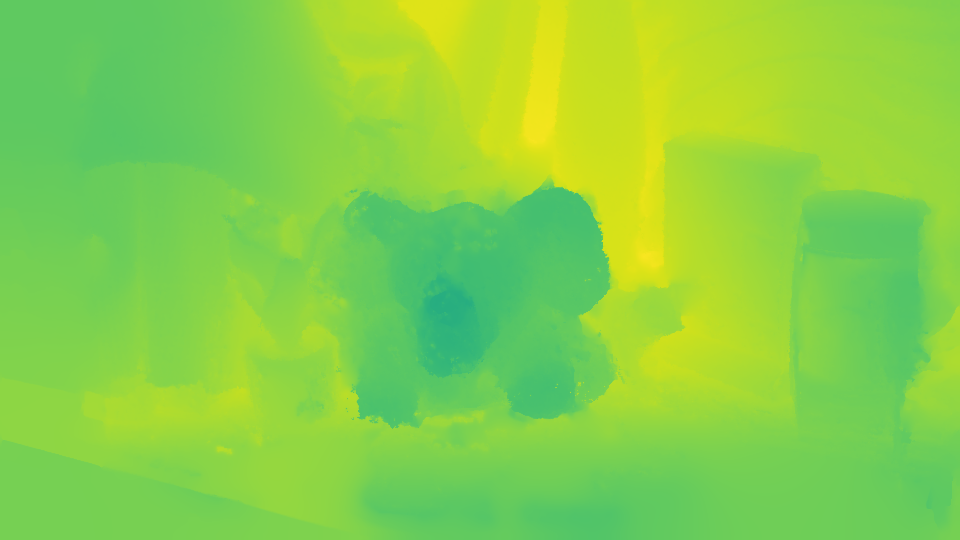}}&
        \raisebox{-.5\height}{\includegraphics[width=\imgw]{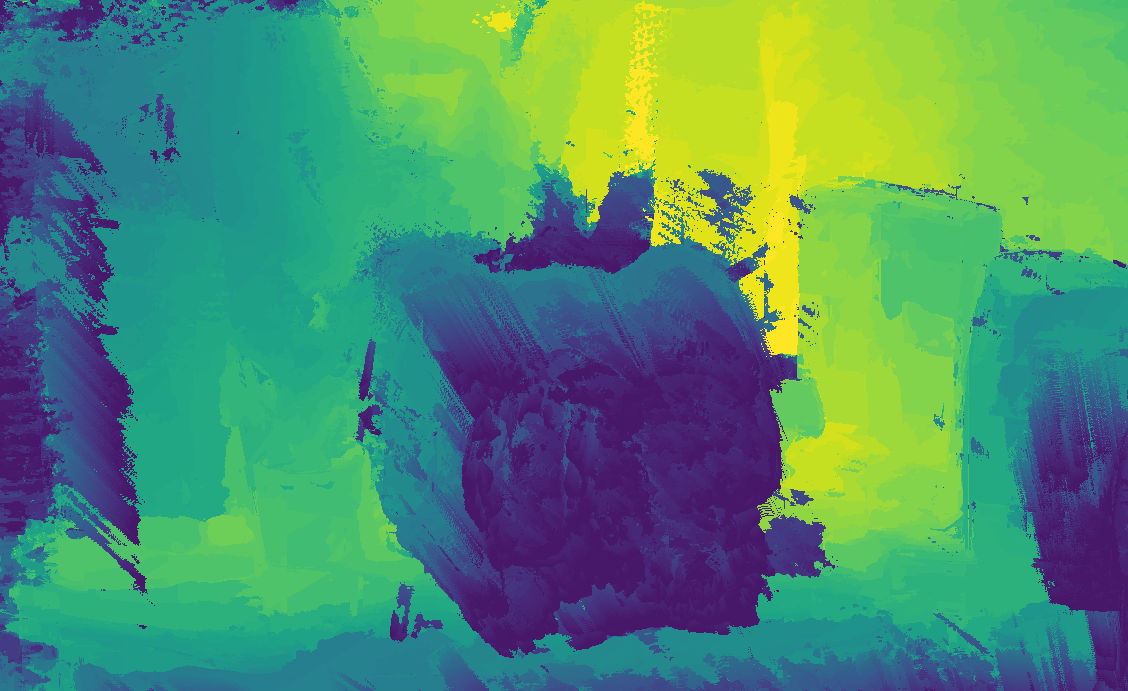}} \\
    
        Ours & 
        EVS \cite{evs} & 
        Ours & 
        EVS \cite{evs} & 
        Ours &
        EVS \cite{evs} \\
    \end{tabular}
    
    \caption{Comparison with Extreme View Synthesis \cite{evs} on the Cat-Dog, Jumping and Elephant-wiggle sequences. Our method produces fewer artifacts than EVS.}
    \label{fig:exp_evs}
\end{figure*}
Figure \ref{fig:exp_evs} shows a comparisons with Extreme View Synthesis (EVS)~\cite{evs}. As input, EVS receives our SfM results. As expected, it exhibits flickering since this method is designed for static scenes and does not enforce temporal consistency. In addition, EVS cannot handle high resolution input because of its intense memory usage; we had to lower the resolution of the input video from 1920$\times$1080 to 1280$\times$720. For the same reason we also could not increase the depth resolution of its scene reconstruction step, which leads inaccurate depth maps and thus severe ghosting in the effected areas.

\textbf{Static---Local Light Field Fusion \protect\cite{llff}. }
\begin{figure}[t]
    \centering
    
    \begin{tabular}{@{}c@{\hspace{1mm}}c@{}}
    \includegraphics[width=0.49\linewidth]{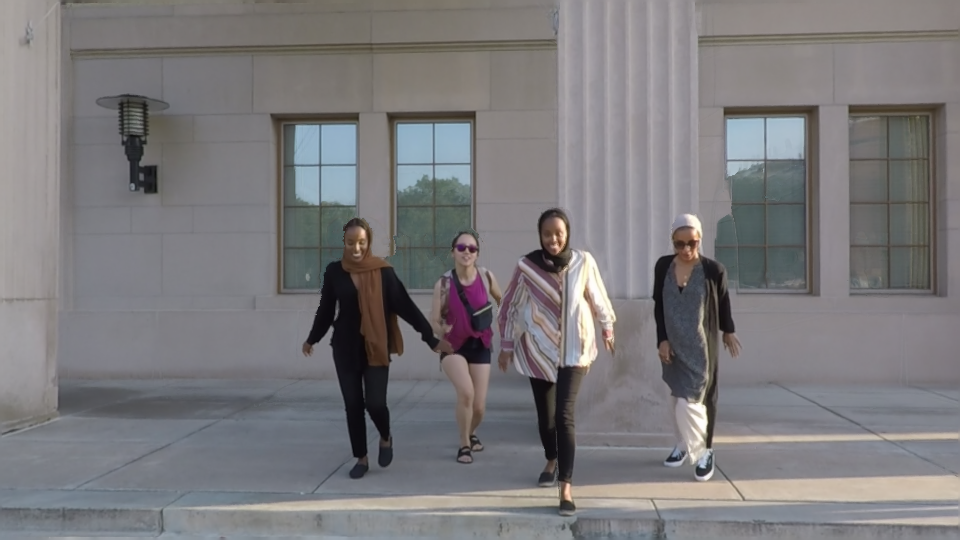} &
    \includegraphics[width=0.49\linewidth]{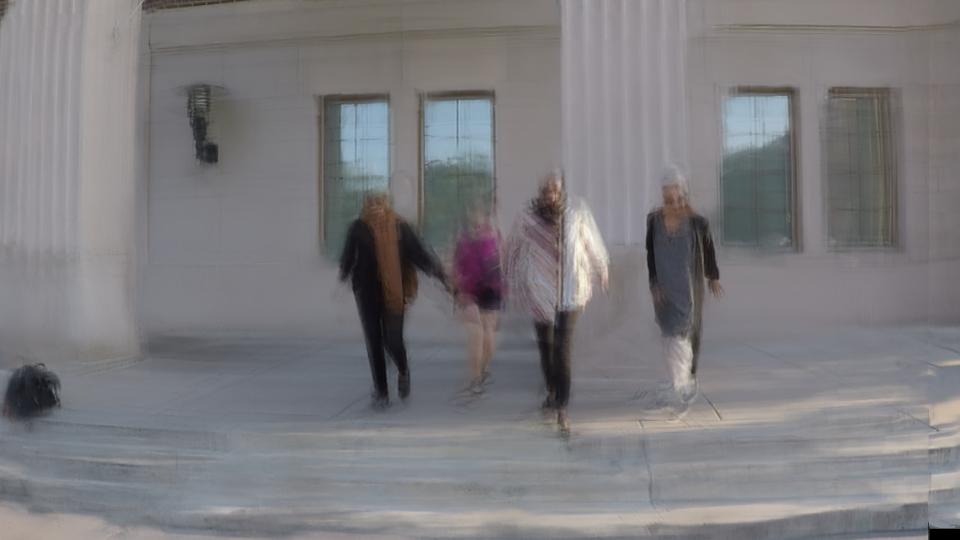} \\
    \includegraphics[width=0.49\linewidth]{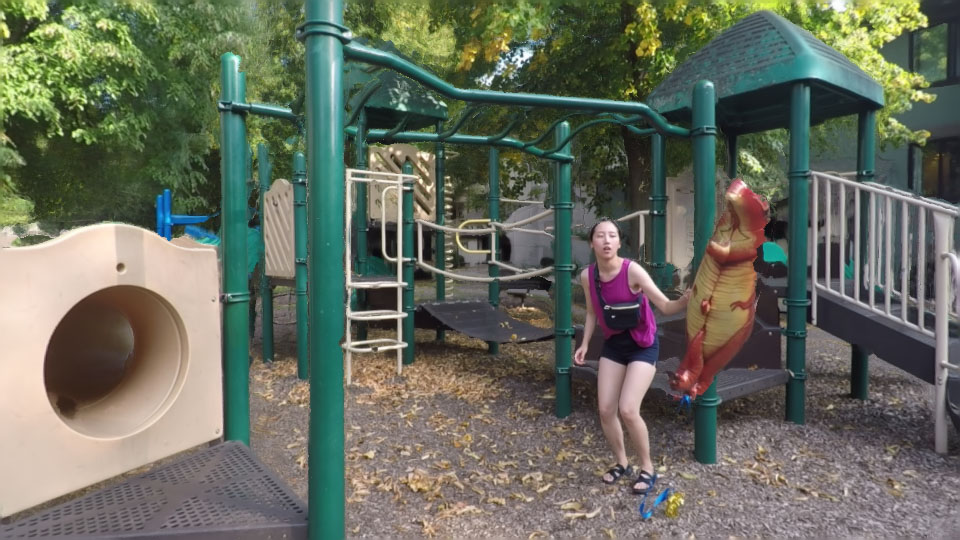} &
    \includegraphics[width=0.49\linewidth]{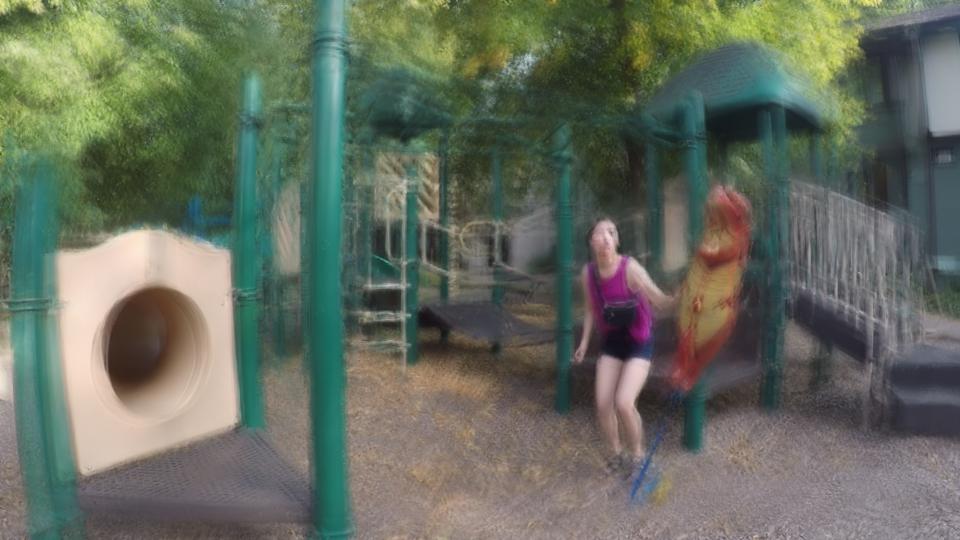}\\

    \includegraphics[width=0.49\linewidth]{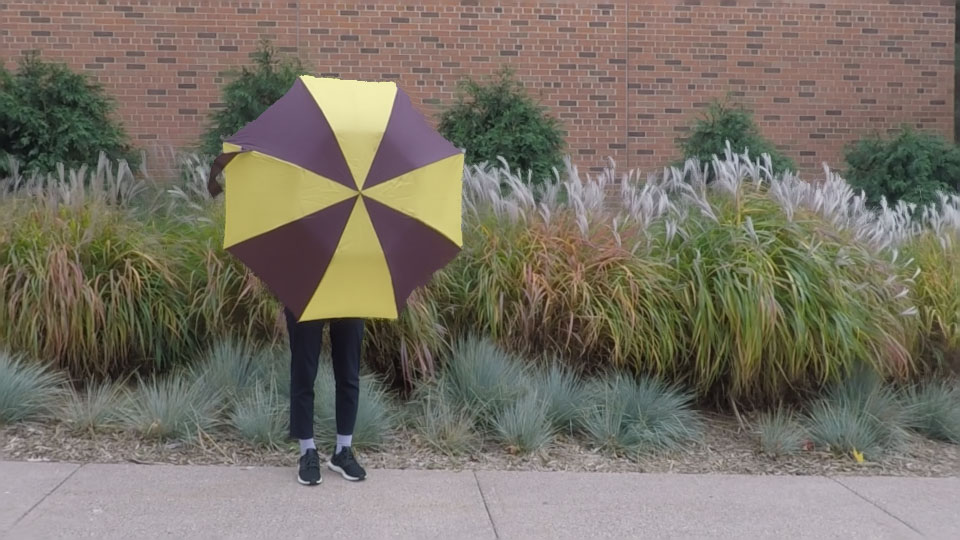} &
    \includegraphics[width=0.49\linewidth]{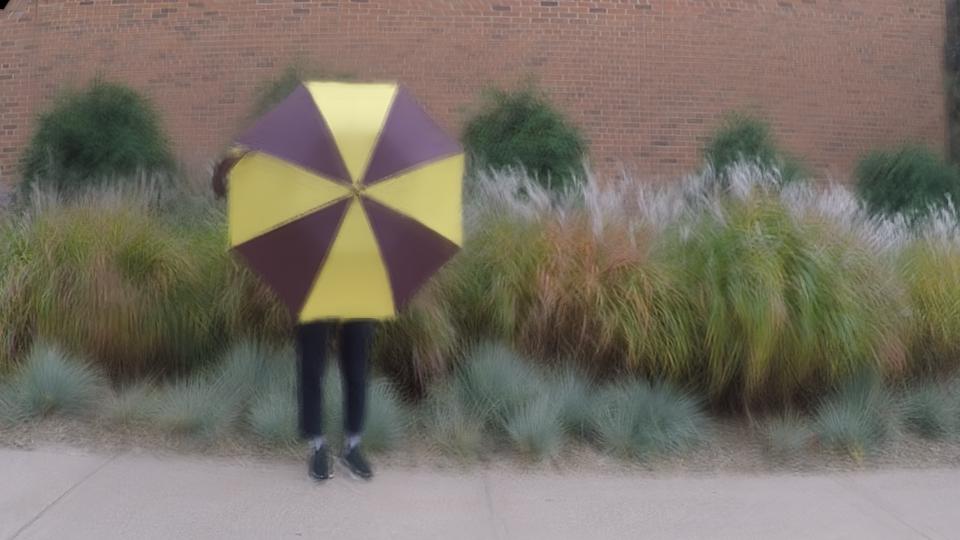}\\
    \includegraphics[width=0.49\linewidth]{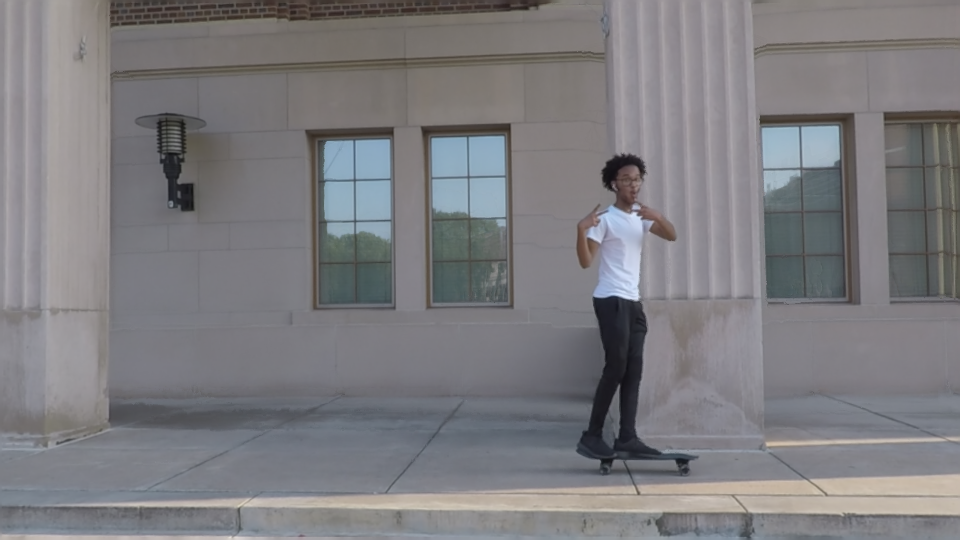} &
    \includegraphics[width=0.49\linewidth]{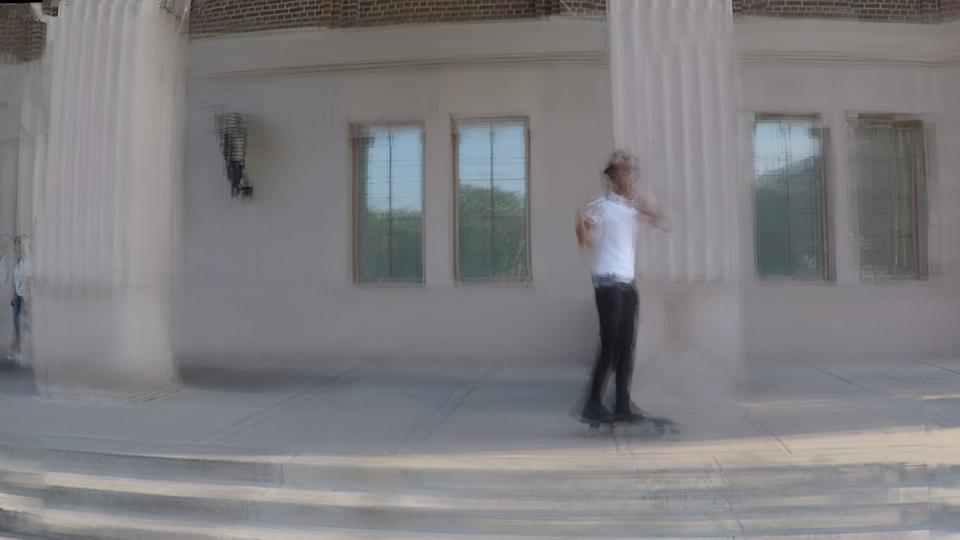} \\
    
    Ours & LLFF\cite{llff}
    \end{tabular}

    \caption{Our method (left) in comparison with Local Light Field Fusion \cite{llff} (right) on the Jumping, Playground, Umbrella and Skating sequences.}
    \label{fig:exp_llff}
\end{figure}
Figure \ref{fig:exp_llff} shows a comparison with Local Light Field Fusion (LLFF)~\cite{llff}. As input, LLFF receives our SfM results. Since our 3D reconstruction is left noisy by design, which is not expected by this method, we fixed the minimum and maximum depths to known correct values LLFF can use for its Multi-Plane Image computations. This reduces the flickering in their video, but it does not eliminate it completely. Our result also appears sharper and with less ghosting artifacts. Since LLFF requires at least 6 cameras to work, we could only compare on the 12-camera dataset sequences.

\textbf{Dynamic---Monocam~\protect\cite{yoon20}. }
\begin{figure*}
    \centering
    
    \begin{tabular}{@{}c@{\hspace{1mm}}c@{\hspace{1mm}}c@{\hspace{1mm}}c@{}}
        \raisebox{-.5\height}{\includegraphics[width=0.22\linewidth]{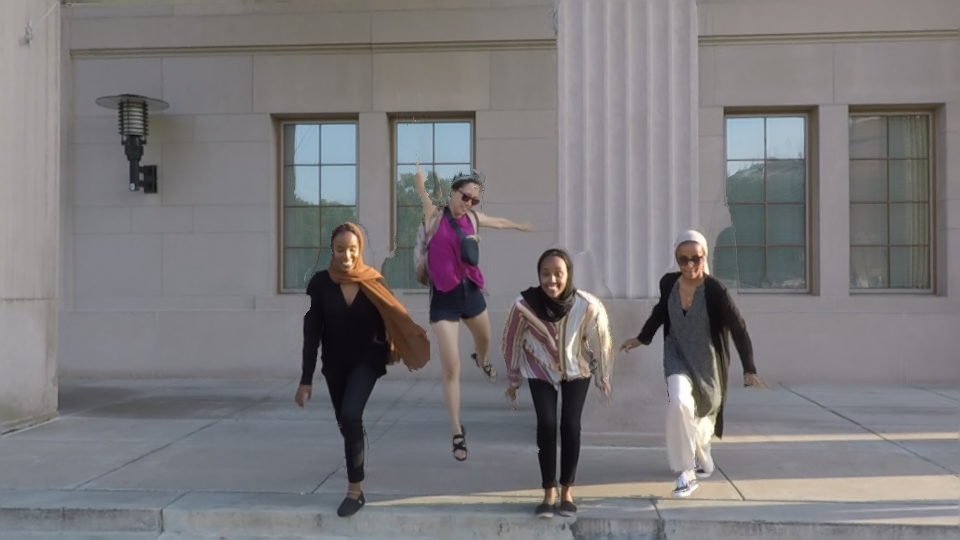}}&
        \raisebox{-.5\height}{\includegraphics[width=0.22\linewidth]{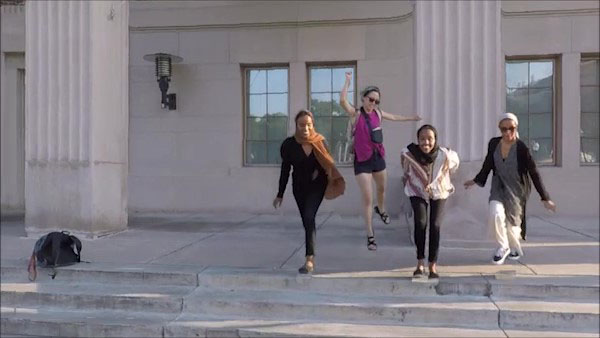}}&
        \raisebox{-.5\height}{\includegraphics[width=0.22\linewidth]{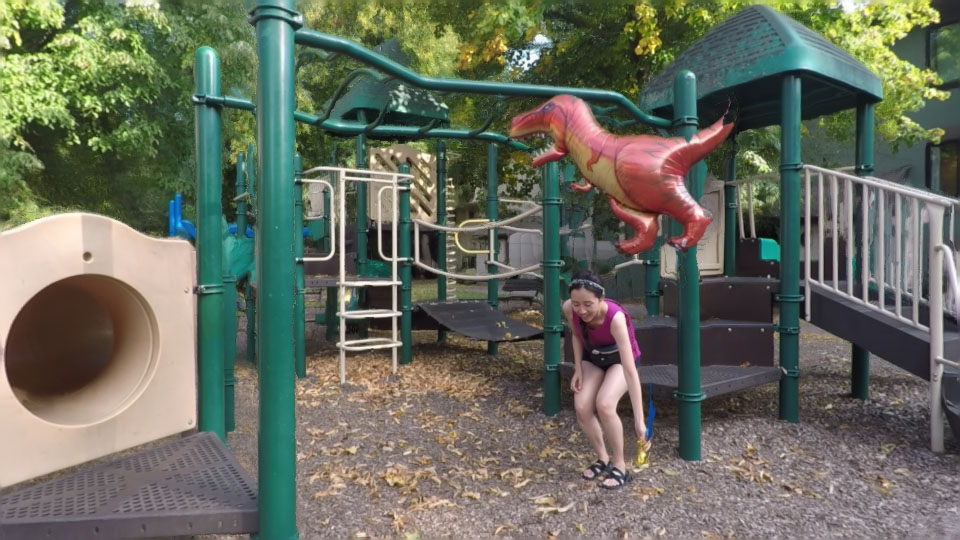}}&
        \raisebox{-.5\height}{\includegraphics[width=0.22\linewidth]{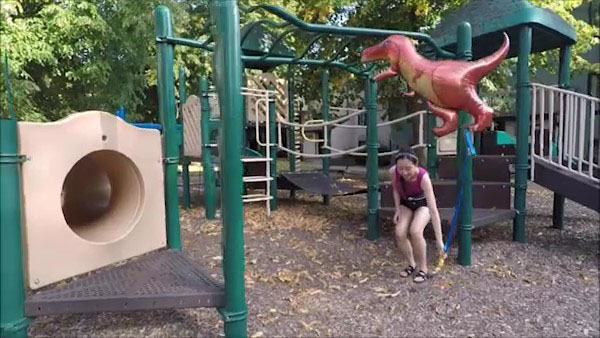}}\\

        \raisebox{-.5\height}{\includegraphics[width=0.22\linewidth]{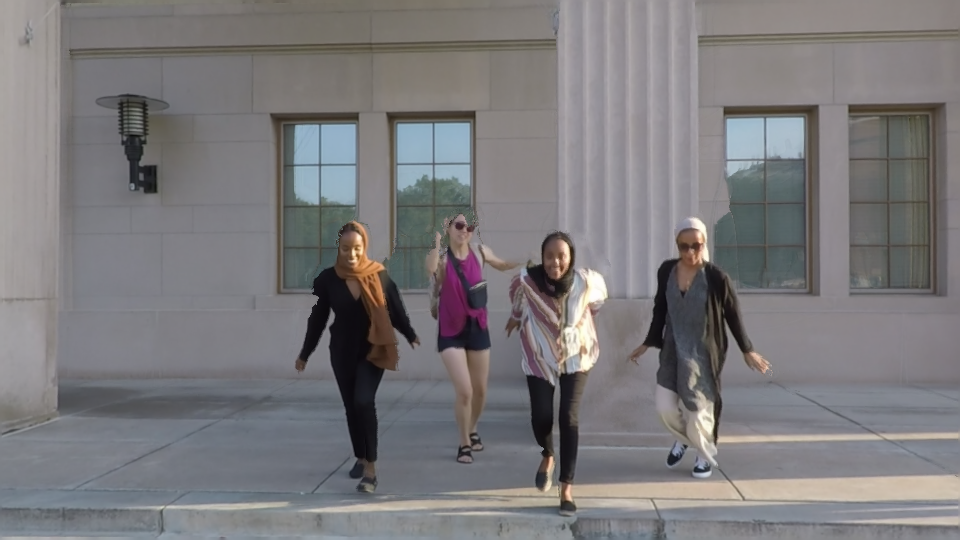}}&
        \raisebox{-.5\height}{\includegraphics[width=0.22\linewidth]{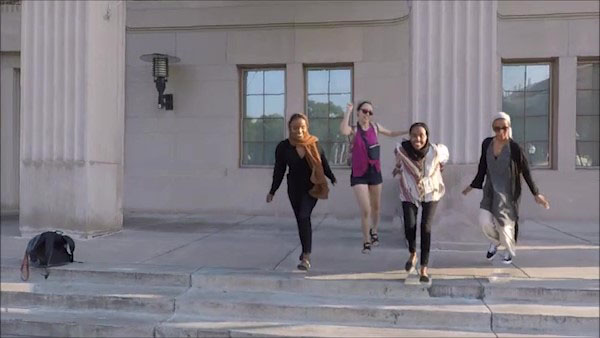}}&
        \raisebox{-.5\height}{\includegraphics[width=0.22\linewidth]{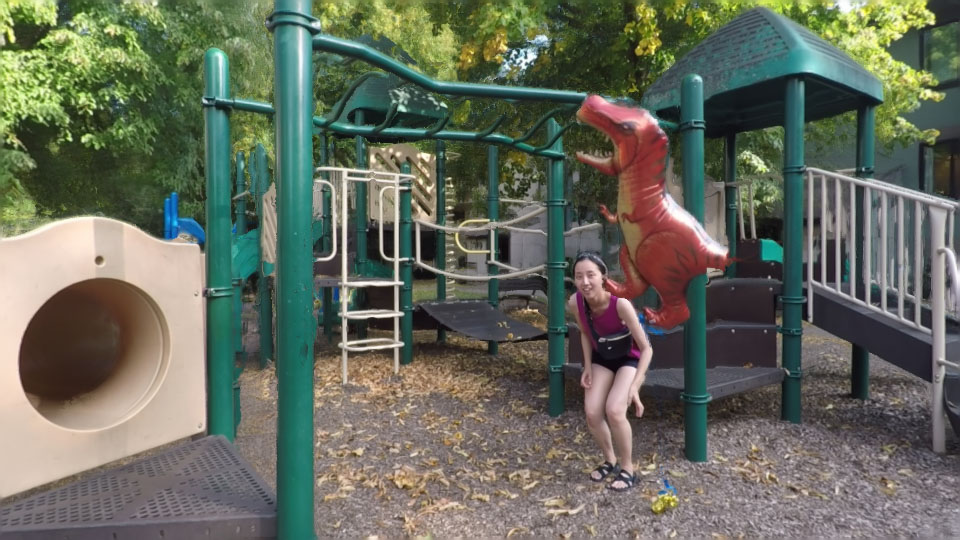}}&
        \raisebox{-.5\height}{\includegraphics[width=0.22\linewidth]{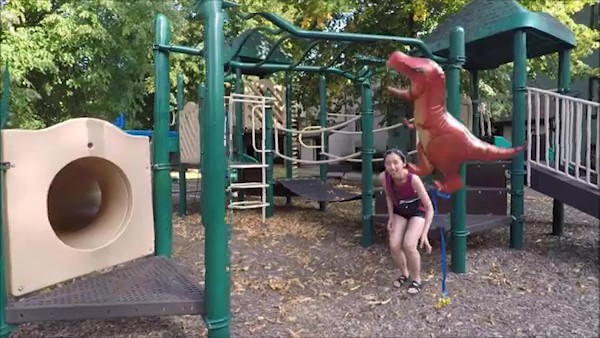}}\\

        \raisebox{-.5\height}{\includegraphics[width=0.22\linewidth]{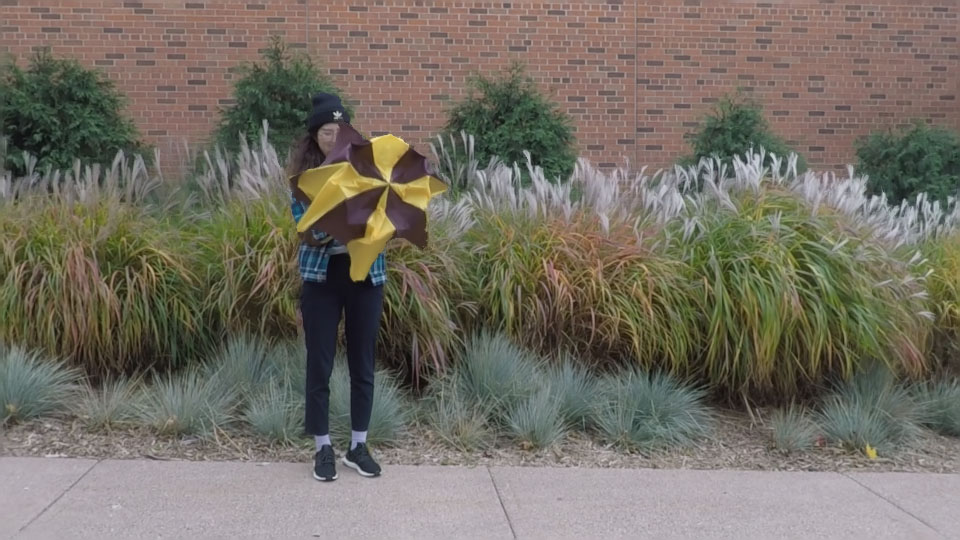}}&
        \raisebox{-.5\height}{\includegraphics[width=0.22\linewidth]{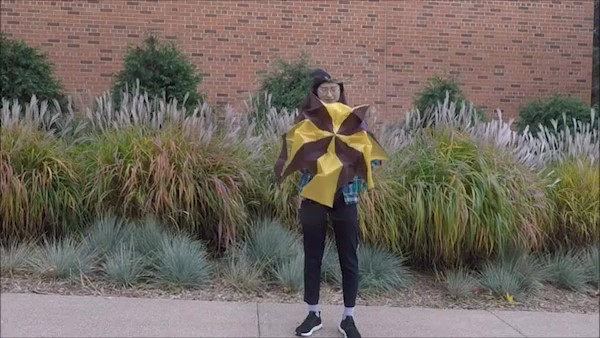}} &
        \raisebox{-.5\height}{\includegraphics[width=0.22\linewidth]{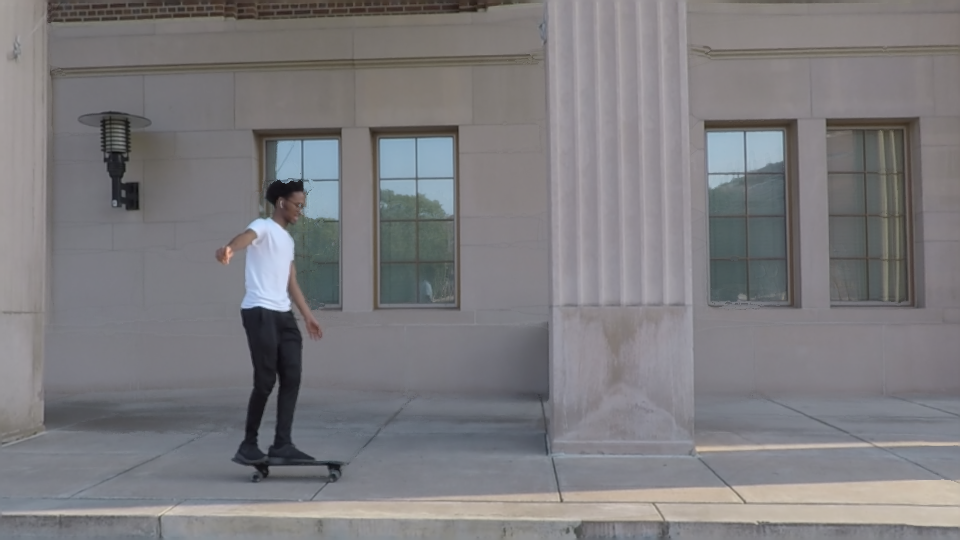}}&
        \raisebox{-.5\height}{\includegraphics[width=0.22\linewidth]{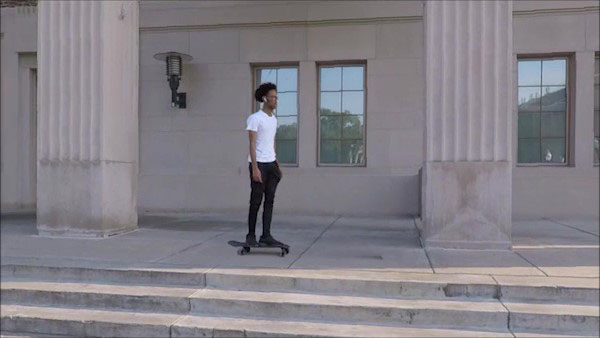}}\\

        \raisebox{-.5\height}{\includegraphics[width=0.22\linewidth]{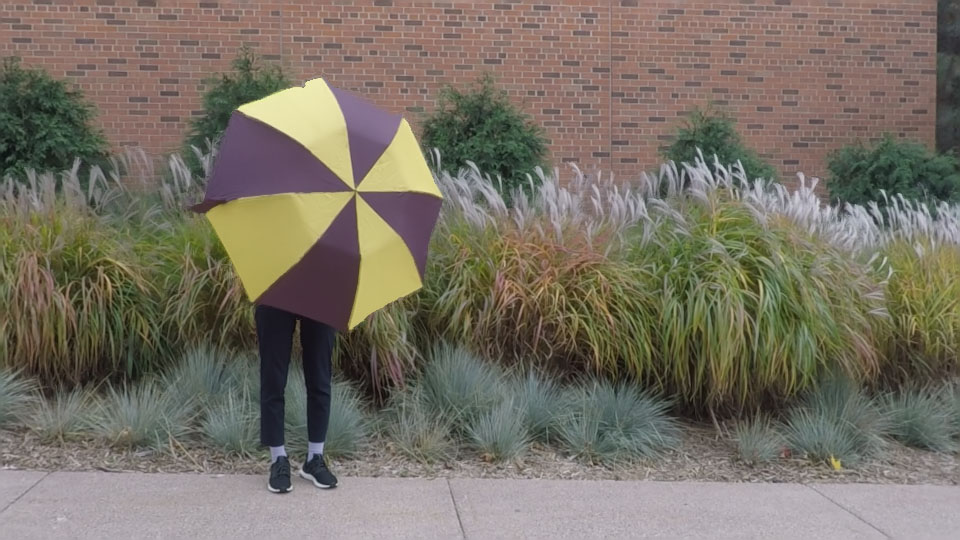}}&
        \raisebox{-.5\height}{\includegraphics[width=0.22\linewidth]{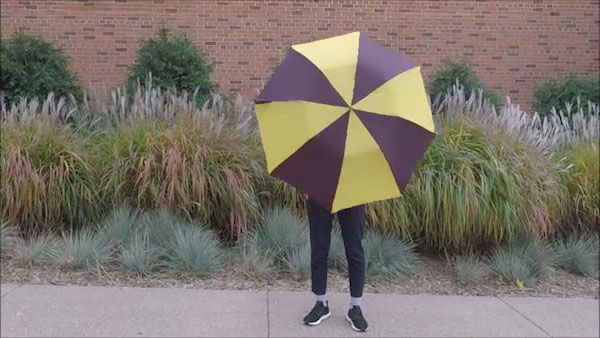}} &
        \raisebox{-.5\height}{\includegraphics[width=0.22\linewidth]{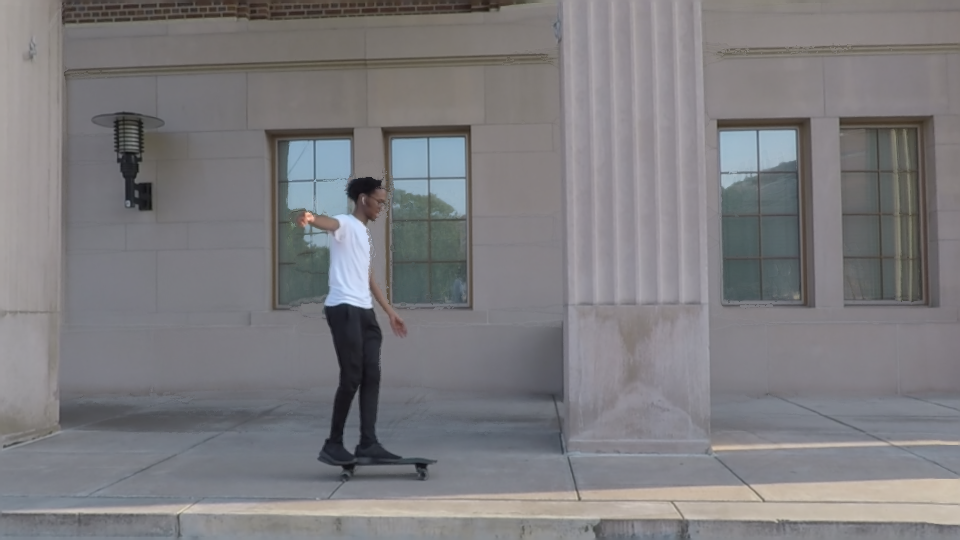}}&
        \raisebox{-.5\height}{\includegraphics[width=0.22\linewidth]{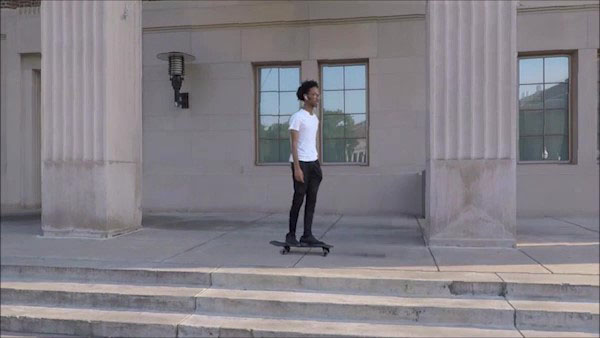}}\\
    
        Ours & 
        Monocam \cite{yoon20} & 
        Ours & 
        Monocam \cite{yoon20} \\
    \end{tabular}
    
    \caption{Comparison with Monocam \cite{yoon20} on the Jumping, Playground, Umbrella and Skating sequences. Stronger temporal inconsistencies in results computed with Monocam can be seen in the accompanying video.}
    \label{fig:exp_mono}
\end{figure*}
Figure \ref{fig:exp_mono} shows a comparison with Monocam~\cite{yoon20}. Here we use the results given by the authors directly for the comparison. It is important to note that the sequences provided by the authors differ slightly from the ones used in the corresponding paper~\cite{yoon20} and for which we have the results. For instance in the skating sequence, the skater is doing hand gestures in the provided input sequence contrary to the published processed result. This nevertheless allows qualitative comparisons.
This figure shows that dynamic background objects like the plants in the umbrella sequence appear static if the virtual camera is static and are not consistent if the virtual camera is dynamic. 
Monocam results also exhibit temporal coherence artifacts.
For instance, the reflections in the jumping and skating sequences jump back and forth based on which view was used to render them. Please see these in the accompanying video.

\textbf{Dynamic---View Interpolation, Virtual Video Camera~\protect\cite{Zitnick2004,Lipski10cgf}. }
\begin{figure}[t]
    \centering
    
    \begin{tabular}{@{}c@{\hspace{1mm}}c@{\hspace{1mm}}c@{}}
        \includegraphics[width=0.32\linewidth,height=2.5cm]{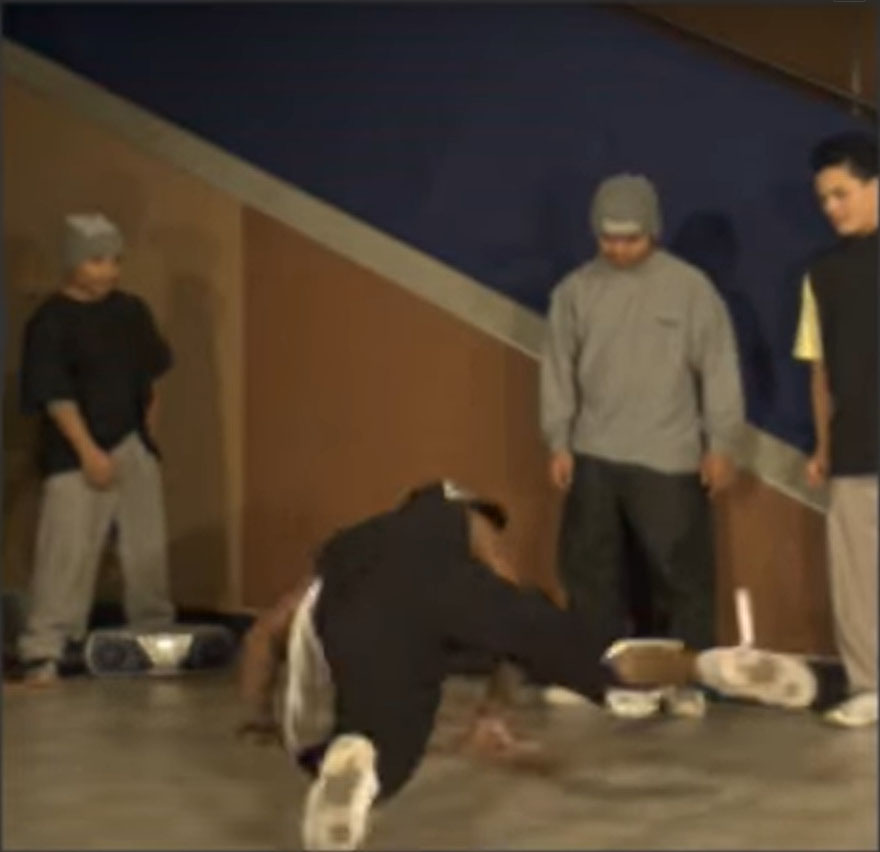}&
        \includegraphics[width=0.32\linewidth,height=2.5cm,]{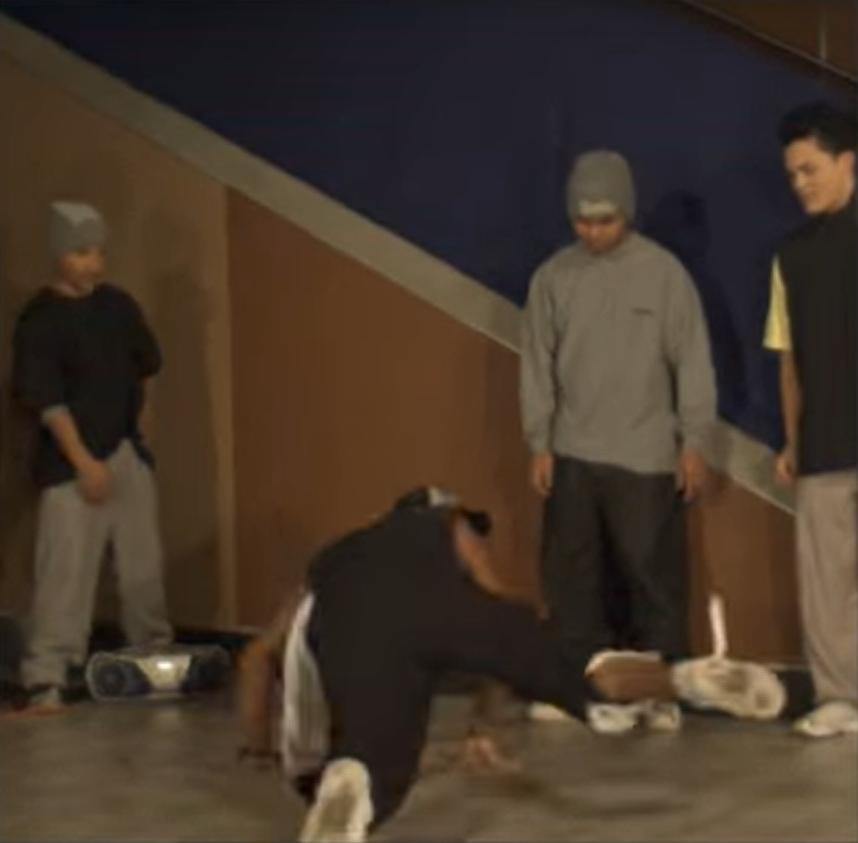}&
        \includegraphics[width=0.32\linewidth,height=2.5cm]{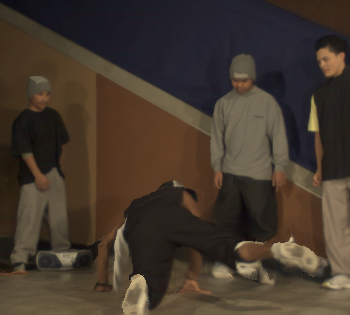}\\
        VI \cite{Zitnick2004} & 
        VCC \cite{Lipski10cgf} &
        Ours \\
    \end{tabular}
    
    \caption{Comparison with \cite{Zitnick2004} and \cite{Lipski10cgf} on the Breakdancers sequence.}
    \label{fig:exp_legacy}
\end{figure}
Figure \ref{fig:exp_legacy} shows a comparison with View Interpolation (VI) \cite{Zitnick2004} and Virtual Video Camera (VVC) \cite{Lipski10cgf} methods. We approximately reproduced the camera path of the video provided by the authors for the Breakdancers scene for this comparison. While VI requires a fixed and calibrated camera grid, our method can handle hand-held devices. VVC eliminates these restrictions, but relies on user input to correct correspondence matches.

\subsection{Challenging Sequence---Drone}

\begin{figure}[t]
    \centering
    \begin{tabular}{@{}c@{\hspace{1mm}}c@{}}
        \includegraphics[width=0.49\linewidth]{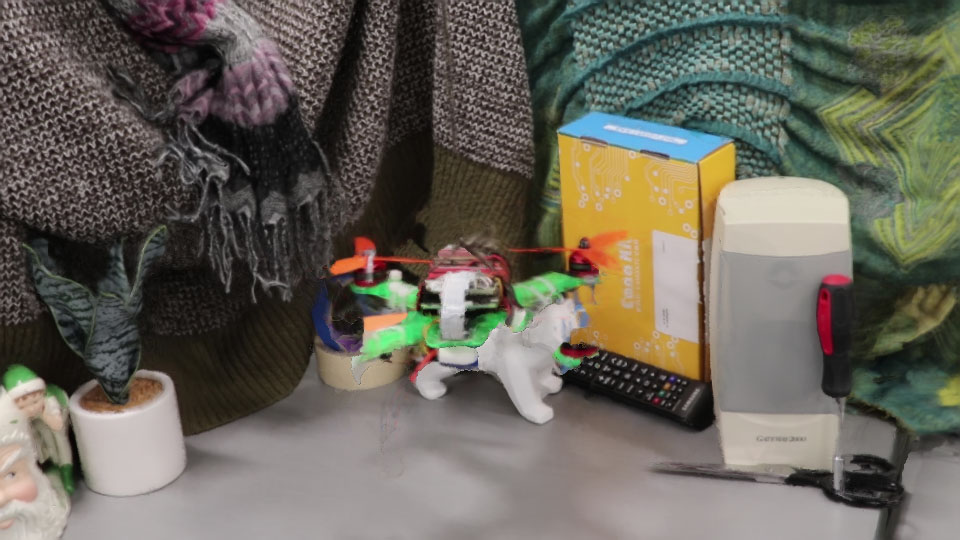}&
        \includegraphics[width=0.49\linewidth]{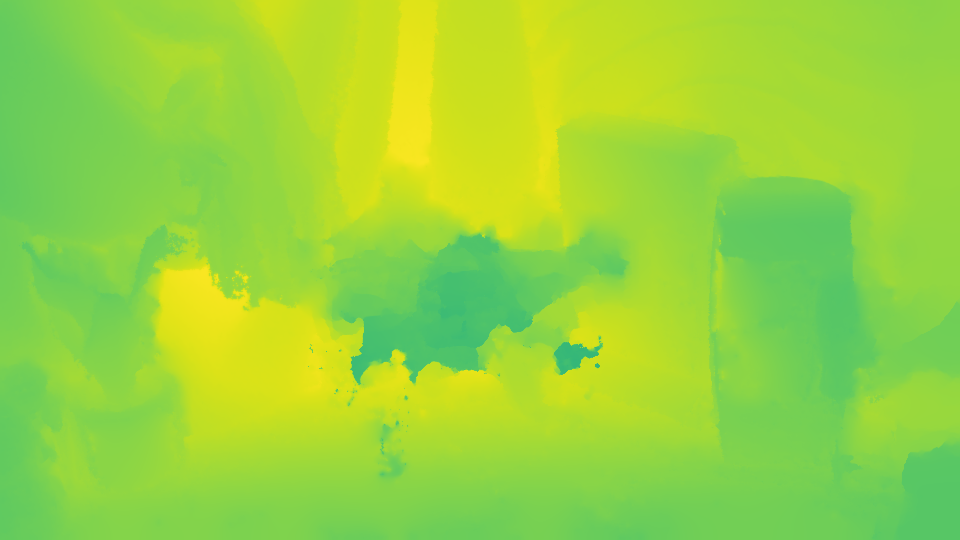}\\
    \end{tabular}
    \caption{\emph{Limitations:} In the drone sequence, the scene has thin features which makes geometry reconstruction difficult. Here, our consistent propagation has trouble correcting for missing sparse feature points.}
    \label{fig:exp_limitation}
\end{figure}

This sequence shows a quadrocopter drone (Figure \ref{fig:exp_limitation}). The drone has many thin features: the chassis, the fan blades, and exposed wires between battery and motors. Here, if our stereo reconstruction fails to find feature points on or nearby thin features at the correct depth, then our consistent propagation cannot provide the correct depth. As such, we see ghosting effects.

\subsection{Computational Resources}
We implemented our system in C++ on a Intel(R) Xeon(R) CPU ES-2630 v3 @2.4GHz computer. We used the OpenMVG library to compute Structure from Motion and sparse depth maps; and both OpenMVS and COLMAP to compute the PatchMatch-based sparse depth map post processing (Section \ref{sec:pose_estimate}).
We parallelize the code using OpenMP and run on 32 cores; the rendering algorithm loads up to 2GBs of data per frame. As an example of wall-clock time, it took 2.2 hours to process the elephant-wiggle sequence ($5$ cameras, $100$ frames per camera). The computation time breaks down to camera and sparse depth estimation (2 hours), and the rendering itself (6.8s per frame, 11.3 minutes for the whole video).

\section{Limitations}

Our method has several limitations. 
First, our choice of the OpenMVG library \cite{moulon2016openmvg} for computing the SfM has the drawback that we must provide focal lengths for a pair of cameras to initiate the reconstruction process.
Second, our method requires that the video sequences should have enough texture on the objects and in the background such that enough SIFT keypoints can be detected and matched.
Another limitation lies in the amount of motion in the frame: conceptually, if SIFT keypoints are only detected on moving objects, then camera pose estimation will fail. In practice, we did not find this to be a problem.
Furthermore, if the baseline is too wide, then not enough points will be obtained on moving objects and the depth propagation will fail.
Finally, our optimization has parameters that can be tuned for each sequences; we provide reasonable initial values (Sec.~\ref{sec:rendering}), but tweaking can improve quality.

Finally, our current implementation is unoptimized C++ running on a CPU. Even if we optimize the implementation, one bottleneck is that keypoints from several images must be matched, and this is time consuming. If we consider SfM as an offline task to be performed once per scene, then the view rendering part currently takes 7 seconds per frame. Given the fixed grid, GPU-based diffusion optimizers are possible, which would produce a much more application-friendly render time.

\section{Conclusion}
We introduce a novel view synthesis method which can handle dynamic scenes. It is based around the key insight that reconstructing temporally-consistent 3D points on dynamic objects is hard, yet a structure-from-motion reconstruction method need not be temporally consistent if temporal consistency can be enforced in the rendering algorithm. We show that this can be accomplished by deferring consistency to a variational screen-space formulation, which makes it easy to robustly enforce spatio-temporal consistency via reprojection constraints weighted by confidences. While our setting has some restrictions, we show competitive results against existing baselines for video-based rendering without using any learning-based approaches. In the future, we hope to reduce constraints in camera motions and temporally with asynchronous videos.

\bibliographystyle{unsrt}
\bibliography{main}

\begin{thebibliography}{10}

\bibitem{vo2016spatiotemporal}
Minh Vo, Srinivasa~G Narasimhan, and Yaser Sheikh.
\newblock Spatiotemporal bundle adjustment for dynamic 3d reconstruction.
\newblock In {\em Proceedings of the IEEE Conference on Computer Vision and
  Pattern Recognition}, pages 1710--1718, 2016.

\bibitem{Mustafa16}
A.~{Mustafa}, H.~{Kim}, J.~{Guillemaut}, and A.~{Hilton}.
\newblock Temporally coherent 4d reconstruction of complex dynamic scenes.
\newblock In {\em 2016 IEEE Conference on Computer Vision and Pattern
  Recognition (CVPR)}, pages 4660--4669, June 2016.

\bibitem{Mustafa19}
Armin Mustafa, Marco Volino, Hansung Kim, Jean{-}Yves Guillemaut, and Adrian
  Hilton.
\newblock Temporally coherent general dynamic scene reconstruction.
\newblock {\em CoRR}, abs/1907.08195, 2019.

\bibitem{holynski2018fast}
Aleksander Holynski and Johannes Kopf.
\newblock Fast depth densification for occlusion-aware augmented reality.
\newblock In {\em SIGGRAPH Asia 2018 Technical Papers}, page 194. ACM, 2018.

\bibitem{zhang2004survey}
Cha Zhang and Tsuhan Chen.
\newblock A survey on image-based rendering—representation, sampling and
  compression.
\newblock {\em Signal Processing: Image Communication}, 19(1):1--28, 2004.

\bibitem{chen1993view}
Shenchang~Eric Chen and Lance Williams.
\newblock View interpolation for image synthesis.
\newblock In {\em Proceedings of the 20th annual conference on Computer
  graphics and interactive techniques}, pages 279--288. ACM, 1993.

\bibitem{gortler1996lumigraph}
Steven~J Gortler, Radek Grzeszczuk, Richard Szeliski, and Michael~F Cohen.
\newblock The lumigraph.
\newblock In {\em Siggraph}, volume~96, pages 43--54, 1996.

\bibitem{flynn2019deepview}
John Flynn, Michael Broxton, Paul Debevec, Matthew DuVall, Graham Fyffe, Ryan
  Overbeck, Noah Snavely, and Richard Tucker.
\newblock Deepview: View synthesis with learned gradient descent.
\newblock In {\em Proceedings of the IEEE Conference on Computer Vision and
  Pattern Recognition}, pages 2367--2376, 2019.

\bibitem{buehler2001unstructured}
Chris Buehler, Michael Bosse, Leonard McMillan, Steven Gortler, and Michael
  Cohen.
\newblock Unstructured lumigraph rendering.
\newblock In {\em Proceedings of the 28th annual conference on Computer
  graphics and interactive techniques}, pages 425--432. ACM, 2001.

\bibitem{shade1998layered}
Jonathan Shade, Steven Gortler, Li-wei He, and Richard Szeliski.
\newblock Layered depth images.
\newblock In {\em Proceedings of the 25th annual conference on Computer
  graphics and interactive techniques}, pages 231--242, 1998.

\bibitem{debevec1996modeling}
Paul~E Debevec, Camillo~J Taylor, and Jitendra Malik.
\newblock Modeling and rendering architecture from photographs: A hybrid
  geometry-and image-based approach.
\newblock In {\em Proceedings of the 23rd annual conference on Computer
  graphics and interactive techniques}, pages 11--20, 1996.

\bibitem{snavely2006photo}
Noah Snavely, Steven~M Seitz, and Richard Szeliski.
\newblock Photo tourism: exploring photo collections in 3d.
\newblock In {\em ACM transactions on graphics (TOG)}, volume~25, pages
  835--846. ACM, 2006.

\bibitem{hedman2016scalable}
Peter Hedman, Tobias Ritschel, George Drettakis, and Gabriel Brostow.
\newblock Scalable inside-out image-based rendering.
\newblock {\em ACM Transactions on Graphics (TOG)}, 35(6):231, 2016.

\bibitem{chaurasia2013depth}
Gaurav Chaurasia, Sylvain Duchene, Olga Sorkine-Hornung, and George Drettakis.
\newblock Depth synthesis and local warps for plausible image-based navigation.
\newblock {\em ACM Transactions on Graphics (TOG)}, 32(3):30, 2013.

\bibitem{achanta2010slic}
Radhakrishna Achanta, Appu Shaji, Kevin Smith, Aurelien Lucchi, Pascal Fua, and
  Sabine S{\"u}sstrunk.
\newblock Slic superpixels.
\newblock Technical report, 2010.

\bibitem{Matzen17}
Kevin Matzen, Michael~F. Cohen, Bryce Evans, Johannes Kopf, and Richard
  Szeliski.
\newblock Low-cost 360 stereo photography and video capture.
\newblock {\em ACM Trans. Graph.}, 36(4), 2017.

\bibitem{Riegler2020FVS}
Gernot Riegler and Vladlen Koltun.
\newblock Free view synthesis.
\newblock In {\em European Conference on Computer Vision}, 2020.

\bibitem{nerf}
Ben Mildenhall, Pratul~P. Srinivasan, Matthew Tancik, Jonathan~T. Barron, Ravi
  Ramamoorthi, and Ren Ng.
\newblock Nerf: Representing scenes as neural radiance fields for view
  synthesis, 2020.

\bibitem{kopf2013image}
Johannes Kopf, Fabian Langguth, Daniel Scharstein, Richard Szeliski, and
  Michael Goesele.
\newblock Image-based rendering in the gradient domain.
\newblock {\em ACM Transactions on Graphics (TOG)}, 32(6):199, 2013.

\bibitem{flynn2016deepstereo}
John Flynn, Ivan Neulander, James Philbin, and Noah Snavely.
\newblock Deepstereo: Learning to predict new views from the world's imagery.
\newblock In {\em Proceedings of the IEEE Conference on Computer Vision and
  Pattern Recognition}, pages 5515--5524, 2016.

\bibitem{zhou2018stereo}
Tinghui Zhou, Richard Tucker, John Flynn, Graham Fyffe, and Noah Snavely.
\newblock Stereo magnification: Learning view synthesis using multiplane
  images.
\newblock {\em arXiv preprint arXiv:1805.09817}, 2018.

\bibitem{llff}
Ben Mildenhall, Pratul~P. Srinivasan, Rodrigo Ortiz-Cayon, Nima~Khademi
  Kalantari, Ravi Ramamoorthi, Ren Ng, and Abhishek Kar.
\newblock Local light field fusion: Practical view synthesis with prescriptive
  sampling guidelines.
\newblock {\em ACM Trans. Graph.}, 38(4), July 2019.

\bibitem{zhou2016view}
Tinghui Zhou, Shubham Tulsiani, Weilun Sun, Jitendra Malik, and Alexei~A Efros.
\newblock View synthesis by appearance flow.
\newblock In {\em European conference on computer vision}, pages 286--301.
  Springer, 2016.

\bibitem{kalantari2016learning}
Nima~Khademi Kalantari, Ting-Chun Wang, and Ravi Ramamoorthi.
\newblock Learning-based view synthesis for light field cameras.
\newblock {\em ACM Transactions on Graphics (TOG)}, 35(6):193, 2016.

\bibitem{deepblending}
Peter Hedman, Julien Philip, True Price, Jan-Michael Frahm, George Drettakis,
  and Gabriel Brostow.
\newblock Deep blending for free-viewpoint image-based rendering.
\newblock {\em ACM Trans. Graph.}, 37(6), December 2018.

\bibitem{evs}
I.~{Choi}, O.~{Gallo}, A.~{Troccoli}, M.~H. {Kim}, and J.~{Kautz}.
\newblock Extreme view synthesis.
\newblock In {\em 2019 IEEE/CVF International Conference on Computer Vision
  (ICCV)}, pages 7780--7789, 2019.

\bibitem{Srinivasan17}
Pratul~P. Srinivasan, Tongzhou Wang, Ashwin Sreelal, Ravi Ramamoorthi, and Ren
  Ng.
\newblock Learning to synthesize a 4d {RGBD} light field from a single image.
\newblock {\em International Conference on Computer Vision (ICCV) 2017}, 2017.

\bibitem{song19}
J.~{Song}, X.~{Chen}, and O.~{Hilliges}.
\newblock Monocular neural image based rendering with continuous view control.
\newblock In {\em ICCV 2019}, pages 4089--4099, 2019.

\bibitem{dos2018navigation}
Rafael~K dos Anjos, Jo{\~a}o Pereira, and Jos{\'e} Gaspar.
\newblock A navigation paradigm driven classification for video-based rendering
  techniques.
\newblock {\em Computers \& Graphics}, 77:205--216, 2018.

\bibitem{Zitnick2004}
C.~Lawrence Zitnick, Sing~Bing Kang, Matthew Uyttendaele, Simon Winder, and
  Richard Szeliski.
\newblock High-quality video view interpolation using a layered representation.
\newblock {\em ACM Trans. Graph.}, 23(3):600–608, August 2004.

\bibitem{Wilburn05}
Bennett Wilburn, Neel Joshi, Vaibhav Vaish, Eino-Ville Talvala, Emilio Antunez,
  Adam Barth, Andrew Adams, Mark Horowitz, and Marc Levoy.
\newblock High performance imaging using large camera arrays.
\newblock {\em ACM Trans. Graph.}, 24(3), 2005.

\bibitem{broxton2020immersive}
Michael Broxton, John Flynn, Ryan Overbeck, Daniel Erickson, Peter Hedman,
  Matthew DuVall, Jason Dourgarian, Jay Busch, Matt Whalen, and Paul Debevec.
\newblock Immersive light field video with a layered mesh representation.
\newblock {\em ACM Transactions on Graphics (Proc. SIGGRAPH)},
  39(4):86:1--86:15, 2020.

\bibitem{relightables}
Kaiwen Guo, Peter Lincoln, Philip Davidson, Jay Busch, Xueming Yu, Matt Whalen,
  Geoff Harvey, Sergio Orts-Escolano, Rohit Pandey, Jason Dourgarian, Danhang
  Tang, Anastasia Tkach, Adarsh Kowdle, Emily Cooper, Mingsong Dou, Sean
  Fanello, Graham Fyffe, Christoph Rhemann, Jonathan Taylor, Paul Debevec, and
  Shahram Izadi.
\newblock The relightables: Volumetric performance capture of humans with
  realistic relighting.
\newblock {\em ACM Trans. Graph.}, 38(6), 2019.

\bibitem{Collet15}
Alvaro Collet, Ming Chuang, Pat Sweeney, Don Gillett, Dennis Evseev, David
  Calabrese, Hugues Hoppe, Adam Kirk, and Steve Sullivan.
\newblock High-quality streamable free-viewpoint video.
\newblock {\em ACM Trans. Graph.}, 34(4), 2015.

\bibitem{Pozo16}
Albert~Parra Pozo, Michael Toksvig, Terry~Filiba Schrager, Joyce Hsu, Uday
  Mathur, Alexander Sorkine-Hornung, Rick Szeliski, and Brian Cabral.
\newblock An integrated 6dof video camera and system design.
\newblock {\em ACM Trans. Graph.}, 38(6), 2019.

\bibitem{penner2017soft}
Eric Penner and Li~Zhang.
\newblock Soft 3d reconstruction for view synthesis.
\newblock {\em ACM Transactions on Graphics (TOG)}, 36(6):235, 2017.

\bibitem{ballan2010unstructured}
Luca Ballan, Gabriel~J Brostow, Jens Puwein, and Marc Pollefeys.
\newblock Unstructured video-based rendering: Interactive exploration of
  casually captured videos.
\newblock {\em ACM Transactions on Graphics (TOG)}, 29(4):87, 2010.

\bibitem{Lipski10cgf}
Christian Lipski, Christian Linz, Kai Berger, Anita Sellent, and Marcus Magnor.
\newblock Virtual video camera: Image-based viewpoint navigation through space
  and time.
\newblock {\em Computer Graphics Forum}, 29(8):2555--2568, Dec 2010.

\bibitem{Luo-VideoDepth-2020}
Xuan Luo, Jia{-}Bin Huang, Richard Szeliski, Kevin Matzen, and Johannes Kopf.
\newblock Consistent video depth estimation.
\newblock {\em ACM Transactions on Graphics (Proceedings of ACM SIGGRAPH)},
  39(4), 2020.

\bibitem{bansal20}
Aayush Bansal, Minh Vo, Yaser Sheikh, Deva Ramanan, and Srinivasa Narasimhan.
\newblock 4d visualization of dynamic events from unconstrained multi-view
  videos.
\newblock {\em The IEEE Conference on Computer Vision and Pattern Recognition
  (CVPR)}, June 2020.

\bibitem{yoon20}
Jae~Shin Yoon, Kihwan Kim, Orazio Gallo, Hyun~Soo Park, and Jan Kautz.
\newblock Novel view synthesis of dynamic scenes with globally coherent depths
  from a monocular camera.
\newblock {\em The IEEE Conference on Computer Vision and Pattern Recognition
  (CVPR)}, June 2020.

\bibitem{bao2019depth}
Wenbo Bao, Wei-Sheng Lai, Chao Ma, Xiaoyun Zhang, Zhiyong Gao, and Ming-Hsuan
  Yang.
\newblock Depth-aware video frame interpolation.
\newblock In {\em Proceedings of the IEEE Conference on Computer Vision and
  Pattern Recognition}, pages 3703--3712, 2019.

\bibitem{davis2003spacetime}
James Davis, Ravi Ramamoorthi, and Szymon Rusinkiewicz.
\newblock Spacetime stereo: A unifying framework for depth from triangulation.
\newblock In {\em 2003 IEEE Computer Society Conference on Computer Vision and
  Pattern Recognition, 2003. Proceedings.}, volume~2, pages II--359. IEEE,
  2003.

\bibitem{moulon2016openmvg}
Pierre Moulon, Pascal Monasse, Romuald Perrot, and Renaud Marlet.
\newblock Openmvg: Open multiple view geometry.
\newblock In {\em International Workshop on Reproducible Research in Pattern
  Recognition}, pages 60--74. Springer, 2016.

\bibitem{Moulon2012}
Pierre Moulon, Pascal Monasse, and Renaud Marlet.
\newblock Adaptive structure from motion with a~contrario model estimation.
\newblock In {\em Proceedings of the Asian Computer Vision Conference (ACCV
  2012)}, pages 257--270. Springer Berlin Heidelberg, 2012.

\bibitem{Cornelis04}
K.~{Cornelis}, F.~{Verbiest}, and L.~{Van Gool}.
\newblock Drift detection and removal for sequential structure from motion
  algorithms.
\newblock {\em IEEE Transactions on Pattern Analysis and Machine Intelligence},
  26(10):1249--1259, Oct 2004.

\bibitem{Barnes09}
Connelly Barnes, Eli Shechtman, Adam Finkelstein, and Dan~B Goldman.
\newblock Patchmatch: A randomized correspondence algorithm for structural
  image editing.
\newblock In {\em ACM SIGGRAPH 2009 Papers}, SIGGRAPH ’09, New York, NY, USA,
  2009. Association for Computing Machinery.

\bibitem{schonberger2016structure}
Johannes~L Schonberger and Jan-Michael Frahm.
\newblock Structure-from-motion revisited.
\newblock In {\em Proceedings of the IEEE conference on computer vision and
  pattern recognition}, pages 4104--4113, 2016.

\end{thebibliography}

\end{document}